\documentclass[12pt]{article}
\pdfoutput=1
\usepackage{amsmath,amsfonts,graphicx,color,bbm,tikz}
\usepackage{comment}
\usepackage[nosort]{cite}
\usepackage{subfigure}
\usetikzlibrary{calc,positioning}
\usetikzlibrary{patterns,arrows,decorations.pathreplacing}
\usepackage{caption}
\usepackage{ulem}
\tikzset{>=stealth}
\usepackage{lipsum}

\newcommand\blfootnote[1]{%
  \begingroup
  \renewcommand\thefootnote{}\footnote{#1}%
  \addtocounter{footnote}{-1}%
  \endgroup}

\makeatletter\@addtoreset{equation}{section}\makeatother
\newcommand{\be}{\begin{equation}}
\newcommand{\ee}{\end{equation}}
\def\beq{\begin{equation}}
\def\eeq{\end{equation}}
\newcommand{\bea}{\begin{eqnarray}}
\newcommand{\eea}{\end{eqnarray}}

\newcommand{\cS}{{\cal S}}

\newcommand{\ket}[1]{{\left| {#1} \right>}}

\renewcommand{\title}[1]{\vbox{\center\LARGE{#1}}\vspace{3mm}}
\renewcommand{\author}[1]{\vbox{\center#1}\vspace{3mm}}

\newcommand{\email}[1]{\vbox{\center\tt#1}\vspace{3mm}}

\begin{document}
\begin{titlepage}
\begin{center}

{\large {\bf Quantum entanglement, supersymmetry, \\ and the generalized Yang-Baxter equation} }

\author{Pramod Padmanabhan,$^a$ Fumihiko Sugino,$^b$ Diego Trancanelli$^{c,\star}$}\blfootnote{${}^\star$ On leave of absence from the Institute of Physics at the University of S\~ao Paulo, S\~ao Paulo, Brazil.}

\vskip -1cm
{$^a${\it Center for Theoretical Physics of Complex Systems,\\ Institute for Basic Science, Daejeon, South Korea}}
\vskip0.1cm
{ $^b${\it Center for Theoretical Physics of the Universe,\\
Institute for Basic Science, Daejeon, South Korea} 
\vskip0.1cm
$^c${\it Dipartimento di Scienze Fisiche, Informatiche e Matematiche, \\
Universit\`a di Modena e Reggio Emilia, via Campi 213/A, 41125 Modena, Italy \\ \& \\
INFN Sezione di Bologna, via Irnerio 46, 40126 Bologna, Italy}}
\email{pramod23phys, fusugino, dtrancan@gmail.com}

\vskip 1cm 
\end{center}

\abstract{
\noindent 
Entangled states, such as the Bell and GHZ states, are generated from separable states using matrices known to satisfy the Yang-Baxter equation and its generalization. This remarkable fact hints at the possibility of using braiding operators as quantum entanglers, and is part of a larger speculated connection between topological and quantum entanglement. We push the analysis of this connection forward, by showing that supersymmetry algebras can be used to construct large families of solutions of the spectral parameter-dependent generalized Yang-Baxter equation. We present a number of explicit examples and outline a general algorithm for arbitrary numbers of qubits. The operators we obtain produce, in turn, all the entangled states in a multi-qubit system classified by the Stochastic Local Operations and Classical Communication protocol introduced in quantum information theory.  
}

\end{titlepage}
\tableofcontents 

\section{Introduction}

Quantum entanglement is one of the most important distinguishing feature between the quantum and the classical worlds. Current technological endeavors to harness the `quantumness' of Nature require an understanding of how to generate and maintain entanglement, protecting the system from decoherence. One effort in this direction is topological quantum computing \cite{Kit1, Nay1, Nay2, Nay3, Nay4}, with anyons 
being proposed as a way to realize fault-tolerant quantum gates thanks to the topological nature of their world-lines. These give rise to {\it braiding operators} solving the {\it Yang-Baxter equation (YBE)} \cite{y, b}, a consistency condition which appears in various contexts including quantum integrable models \cite{korebook} and knot theory \cite{lhbook}.\footnote{For a historical introduction to the YBE see \cite{perkybe}.}

In the last decade, a concrete relation between braiding operators and quantum information theory has been proposed in a series of papers by Kauffman, Lomonaco and collaborators, see {\it e.g.} \cite{lh1,lh2, lh3, lh4, lh5}, where it  was shown that the Bell matrix, the two-qubit gate that produces the maximally entangled Bell states out of separable states, solves the YBE. This remarkable fact suggests that entangling gates may be thought as braiding operators, establishing a deep connection between quantum and topological entanglement.\footnote{Note that all the entangling gates 
cannot be interpreted as braiding operators, as a random matrix generates entanglement but hardly satisfies the YBE~\cite{collins}. It would be interesting to consider which kinds of 
entangling gates are also braiding operators.}
  A precursor example of such connection had already been put forward in \cite{aravind}, relating the GHZ state of three qubits to Borromean rings: both are maximally entangled systems of three components, which become completely unentangled upon the removal of one of the components.\footnote{This early example has to be taken with a grain of salt, since it relies on the choice of a particular basis for the three-qubit system. See also \cite{sugita,Quinta:2018scm}.} Subsequently, it was shown in \cite{alga} that quantum entanglement is necessary to detect topological entanglement: if the solution to the YBE is non-entangling, the corresponding link invariant is incapable of distinguishing topologically different knots. From the point of view of quantum computing, it was also shown that entangling operators are necessary to form a universal set of gates \cite{bry}.

A natural question that arises is whether generic entangled states in multi-qubit systems can also be produced from solutions to the YBE, the so-called {\it $R$-matrices}. Besides the already mentioned Bell matrix, this was shown to be the case for the GHZ states in \cite{r1,r2}. As the $R$-matrices producing the GHZ states must act on three qubits simultaneously, this necessitates the introduction of the {\it generalized Yang-Baxter equation (gYBE)}, which accommodates $R$-matrices with support on more than two qubits. Since then, solutions of the gYBE have been constructed in \cite{kitwan} using fusion ribbon category and a complete classification of the solutions of \cite{r1,r2} has been found in \cite{rchen}. Such solutions are built from the generators of {\it extraspecial 2-groups}, which were further studied in \cite{z1}. A Majorana fermion realization of extraspecial 2-group generators appeared in \cite{lhm}, while multi-qudit generalizations were considered in \cite{r4, c1, ho}.

To push this connection between topological and quantum entanglement further, it is essential to generate new entangled states from $R$-matrices, in addition to the states mentioned above, namely the Bell and GHZ states. To this scope, it is useful to consider the {\it  Stochastic Local Operations and Classical Communication (SLOCC)} protocol of quantum information theory, since this provides a classification of the different ways in which a multi-qubit system can be entangled \cite{Cirac}. More specifically, in a three-qubit system there are two inequivalent, maximally entangled classes of states, the GHZ states and the so-called {\it W-states}. Both the GHZ and W-classes generalize to the multi-qubit sectors, whose full classification under the SLOCC protocol is however unknown \cite{Cirac}. Another class which appears in every multi-qubit system is the class of partially entangled states made up of GHZ and W-state classes. 

It is then natural to ask if there exist (unitary) $R$-matrices that generate W-states as well as partially entangled states, or, more in general, representative states of all different classes of the SLOCC classification. We answer this question in the affirmative by explicitly constructing such $R$-matrices for systems of arbitrary numbers of qubits. We achieve this by using certain supersymmetry algebras in quantum mechanics as a solution-generating technique. Starting from certain Ans\"atze for the $R$-matrices, we are able to generate large families of solutions of the gYBE and to show that for every multi-qubit entangled state in a SLOCC class there is a canonical way to construct the unitary $R$-matrices that solve the gYBE and generate that state. This extends very naturally to qudits using para-supersymmetry which we will cover in a companion paper. Moreover, states in different SLOCC classes are seen to be related in a very simple way by the action of the supersymmetry generators, the {\it supercharges}.

This paper is organized as follows. We start with a brief review of the YBE in Sec.~\ref{ybeGybe}, followed by the definition of the gYBE. Solutions of the YBE in terms of permutation operators, commuting projectors and nilpotent operators are also discussed. These will serve as Ans\"atze for the new solutions we find. Next we review the supersymmetry algebra in $0+1$ dimension in Sec.~\ref{susyAlgebra} and a local realization thereof in terms of {\it Symmetric Inverse Semigroups (SISs)}. While a plethora of different solutions, both unitary and non-unitary, are obtainable with our method, we will focus on those that provide the different types of entangled states under the SLOCC protocol described in Sec.~\ref{SLOCCclasses}. We study the two-qubit and the three-qubit cases in detail, whereas for the multi-qubit case we restrict our attention to the SLOCC classes that can be generalized from the three-qubit sector. The $R$-matrices producing these states are obtained from supersymmetry, as detailed in Sec.~\ref{susyR}, which contains the bulk of our results. In that section we also show how to recover the unitary solutions of \cite{r1,r2} in our general construction. We conclude with comments about the relation between the solutions we find and certain braid-like algebras and suggest future directions in Sec.~\ref{remarks}. A short appendix introduces a non-trivial supercharge that relates two different SLOCC classes.

\section{The Yang-Baxter equation and its generalization}
\label{ybeGybe}

Consider the Hilbert space $\mathcal{H}_\textrm{total} = \otimes_{i=1}^N~\mathcal{H}_i$ of a composite system of local Hilbert spaces $\mathcal{H}_i$ at sites $i=1,\ldots, N$. 

The YBE is an operator equation for an invertible matrix $R$. It can be formulated in various equivalent ways, the one of most interest to us being the case in which the $R$-matrix acts on two consecutive sites, $R_i:\mathcal{H}_i\otimes \mathcal{H}_{i+1} \rightarrow \mathcal{H}_i\otimes \mathcal{H}_{i+1}$, and the YBE is given by
\begin{equation}
\label{rbraid}
R_i(u)R_{i+1}(u+v)R_i(v) = R_{i+1}(v)R_i(u+v)R_{i+1}(u),
\end{equation}
or, writing  $R_i(u)\equiv R_{i\, i+1}(u)$, by 
\begin{equation}
\label{rbraid2}
R_{i\, i+1}(u)R_{i+1\, i+2}(u+v)R_{i\, i+1}(v) = R_{i+1\, i+2}(v)R_{i\, i+1}(u+v)R_{i+1\, i+2}(u).
\end{equation}
The representation \eqref{rbraid} depends on a {\it spectral parameter}, $u$, and is known as the {\it braided form} of the YBE,\footnote{\label{foot3} The most general form of the $R$-matrix depends on two spectral parameters as 
$R_i(u,v)$, and satisfies 
\begin{equation}
R_i(u,v)R_{i+1}(u,w)R_i(v,w) = R_{i+1}(v,w)R_i(u,w)R_{i+1}(u,v).\nonumber
\end{equation}
By assuming that $R_i(u,v)$ depends on $u$ and $v$ only through the difference $u-v$, one recovers (\ref{rbraid}). 
}  
as it resembles the relation satisfied by the generators $\sigma_i$ of the {\it braid group}
\begin{equation}
\label{braid}
\sigma_i\sigma_{i+1}\sigma_i = \sigma_{i+1} \sigma_i \sigma_{i+1}.	
\end{equation}
The main difference between the YBE and the braid relation above is the dependence of the former on the spectral parameter. The process of obtaining a solution of the YBE using the generators of the braid group is known as {\it Baxterization} \cite{jonesBaxterization}. Besides \eqref{braid}, the braid group generators must also satisfy the {\it far-commutativity} condition
\begin{equation}
\sigma_i\sigma_j = \sigma_j\sigma_i, \qquad |i-j|>1.
\end{equation}

In the integrability literature, the YBE is usually quoted in a different form, which can however be easily shown to be equivalent to \eqref{rbraid} or \eqref{rbraid2}:
\begin{equation}
\label{rcheck}
\check{R}_{12}(u) \check{R}_{13}(u+v) \check{R}_{23}(v)= \check{R}_{23}(v) \check{R}_{13}(u+v) \check{R}_{12}(u).
\end{equation}
Here $\check{R}_{ij}=R_{ij}P_{ij}$, with $R_{ij}$ being the $R$-matrix above, now acting trivially on all sites except the $i$-th and $j$-th, which need not be consecutive, and $P_{ij}$ being the permutation operator, $ P_{i j} : \mathcal{H}_i\otimes \mathcal{H}_j \rightarrow \mathcal{H}_j\otimes \mathcal{H}_i.$ A crucial property of the permutation operator is that $P_{i j}^2= I$, making it invertible with the inverse equal to itself. Furthermore, this operator satisfies
\begin{equation}
\label{perm}
P_{i\, i+1} P_{i+1\, i+2} P_{i\, i+1} = P_{i+1\, i+2} P_{i\, i+1} P_{i+1\, i+2},	
\end{equation}
which is identical to \eqref{braid} upon identifying $\sigma_i \sim P_{i\, i+1}$. The difference between the braid group  and the permutation group generated by $P_{i\,i+1}$ is that $\sigma_i^{-1}\neq \sigma_i$.

A simple solution to the YBE in \eqref{rbraid} can be found by precisely exploiting \eqref{perm} and starting from the simple Ansatz
\begin{equation}
\label{ransatz}
R_{i}(u) = I + a(u)P_{i\,i+1}	,
\end{equation}
with $a(u)$ an unknown function of the spectral parameter $u$. Substituting into \eqref{rbraid} and equating coefficients gives a simple functional equation for $a(u)$
\begin{equation}
a(u)+a(v) = a(u+v),	
\end{equation}
which is solved by $a(u)=c u$ for some constant $c$.

A second solution is provided by projectors $e_i^2=ke_i$ satisfying  $e_i e_{i+1} e_i = e_{i+1} e_i e_{i+1}$,
with $e_i$ supported on the consecutive sites $i$ and $i+1$, and $k$ a normalization factor. 
A simple choice of operators satisfying the braid relation are commuting projectors ($e_ie_j=e_je_i$). Using a similar Ansatz to \eqref{ransatz},
\begin{equation}
\label{Rcp}
R_i(u) = I +a(u)e_i,	
\end{equation}
one finds that $a(u)$ must obey 
\bea
a(u) + a(v) + ka(u) a(v) = a(u+v),
\eea
which is solved by $a(u) = \left(e^{c u}-1\right)/k$ for some constant $c$.

Finally, another solution can be found by considering nilpotent operators $Q_i$, such that $Q_i^2=0$ and obeying $Q_iQ_{i+1}Q_i = Q_{i+1}Q_iQ_{i+1}$. With the Ansatz
\begin{equation}
R_i(u) = I + a(u)Q_i,
\label{Rqi}
\end{equation}
one finds again that $a(u)$ must be linear in $u$, as in \eqref{ransatz}. This operator is invertible with the inverse given by $R_i(-u)$. Unlike \eqref{ransatz} and \eqref{Rcp}, this solution is non-unitary as the nilpotent operator is non-hermitian. 

In order to generalize the braided form of the YBE to more general situations, it is useful to rewrite \eqref{rbraid} as 
\begin{equation}
\left(R\otimes I\right) \left(I\otimes R\right) \left(R\otimes I\right) = \left(I\otimes R\right) \left(R\otimes I\right) \left(I\otimes R\right).
\end{equation}
This is also known as the $(d,2,1)$-YBE, where $d$ denotes the dimension of the local Hilbert space, ``2" denotes the number of copies of local Hilbert spaces on which the $R$-matrix acts, and ``1" denotes the number of copies of the identity $I$ appearing in each parenthesis of the YBE. This naturally generalizes to the so-called $(d,m,l)$-gYBE, which is written as 
\begin{equation}\label{rgybe}
\left(R\otimes I^{\otimes l}\right) \left(I ^{\otimes l}\otimes R\right) \left(R\otimes I ^{\otimes l}\right) = \left(I ^{\otimes l}\otimes R\right) \left(R\otimes I ^{\otimes l}\right) \left(I ^{\otimes l}\otimes R\right),	
\end{equation}
with the $R$-matrix now acting on $m$ consecutive copies of the local $d$-dimensional Hilbert spaces, {\it i.e.} $R:\mathcal{H} ^{\otimes m}\rightarrow \mathcal{H} ^{\otimes m}$. The parameter $l$ is introduced so as to ensure that the generalized $R$-matrices satisfy far-commutativity, which is needed to ensure that they can be used to construct representations of the braid group. 

The form of the $(d,m,l)$-gYBE with the spectral parameter dependence is given by
\begin{equation}
\label{gYBEparameters}
\left(R(u)\otimes I^{\otimes l}\right) \left(I ^{\otimes l}\otimes R(u+v)\right) \left(R(v)\otimes I ^{\otimes l}\right) = \left(I ^{\otimes l}\otimes R(v)\right) \left(R(u+v)\otimes I ^{\otimes l}\right) \left(I ^{\otimes l}\otimes R(u)\right).
\end{equation}

\section{Supersymmetry in $0+1$ dimension}
\label{susyAlgebra}

At the heart of the solution-generating technique proposed in this paper lies the idea of $\mathbb{Z}_2$-graded Hilbert spaces, explicitly realized in our setup by supersymmetry in $0+1$ dimension. The supersymmetry algebra is generated by a nilpotent operator -- a {\it supercharge} -- $q$ and its adjoint $q^\dag$, which map the `bosonic' and `fermionic' sectors of the Hilbert space into one another. The supercharges satisfy
\begin{equation}
\label{susyrelations}
q^2=\left(q^\dag\right)^2=0,\qquad \{q, q^\dag\}=h,
\end{equation}
where $h$ is a Hamiltonian. It follows from this algebra that $[h,q]=[h,q^\dag]=0$, so that $h$ is supersymmetric. We can think of the Hamiltonian as the sum of two operators $b\equiv q q^\dag$ and $f\equiv q^\dag q$, which project onto the bosonic and fermionic parts of the Hilbert space, respectively. In fact, $b$ and $f$ are orthogonal to each other, as can be easily verified from \eqref{susyrelations}.  Note, however, that these are not fully fledged projectors, as $h=b+f\neq 1$ in general, as we shall see below.

In this paper we consider a special kind of supersymmetry, which arises naturally from SISs we shall employ to generate our solutions. In the cases we consider, the Hamiltonian and its bosonic and fermionic parts are idempotent and satisfy
\begin{eqnarray}
\label{sissusy}
& h^2  = h,\qquad b^2=b,\qquad f^2=f,  \cr
& bq= q, \qquad qf=q, \qquad q^\dag b =q^\dag,\qquad fq^\dag= q^\dag. 
\end{eqnarray}
These relations also imply that $hq=qh=q$ and $q^\dag h=hq^\dag = q^\dag$. It is important to emphasize that \eqref{sissusy} need not hold for generic supersymmetric systems, but they do apply to supersymmetric charges built out of SISs and are crucial for constructing our $R$-matrices, as we shall see in Sec.~\ref{susyR}.

Another important ingredient in our construction is a {\it grading operator} $w$, that satisfies
\begin{equation}
\label{witten1}
w^2=1,\qquad  \{q, w\}=\{q^\dag, w\}=0,
\end{equation}
which imply that $[h,w]=0$. This grading operator is also known as the {\it Witten operator} and it is useful for computing the Witten index of the theory under consideration, to check whether supersymmetry is spontaneously broken or preserved \cite{WittenSUSY, SUSYQMreview}. The Witten operator can be explicitly realized as 
\begin{equation}\label{witten2}
w = \left(-1\right)^b = e^{i\pi b} = 1-2b.
\end{equation}
It is easy to check that $w$ satisfies \eqref{witten1}. We could have equivalently used the projector to the fermionic sector $f$, instead of $b$. One can also verify that
\begin{eqnarray}
\label{witten3}
wq=-q,\qquad qw=q, \qquad q^\dag w=-q^\dag,\qquad wq^\dag = q^\dag.
\end{eqnarray}

The supersymmetry algebra in \eqref{susyrelations} can be implemented both locally (on a single site) and non-locally (on two or more sites). The local implementation can be obtained by using 
SISs, 
the focus of this paper, whereas the non-local implementation exploits partition algebras and is left for a companion paper. 

\subsection{Local realization via inverse semigroups}

We start with a brief review of SISs, see \cite{sis,ppsusy} for more details. Let $S^n=\{ 1, 2,\ldots, n\}$ and consider the set of all partial bijections on $S^n$ together with the usual  composition rule, which is binary and associative. This pair forms an SIS, denoted by $\cS^n = (S^n, *)$. Consider the set of partial bijections on the subset of $S^n$ of order $p\leq n$ and denote the resulting SIS as $\cS^n_p$. We show the algebra of partial bijections on the subset in a diagrammatic way, by means of a few examples.

\begin{figure}[ht!]
\centering
\begin{tikzpicture}[scale=.6]
    \node (E) at (0,0) {$\bullet$};
    \node[right=of E] (F) {$\bullet$};
    \node[below=of F] (A) {$\bullet$};
    \node[below=of E] (B) {$\bullet$};
    \draw[->,ultra thick] (E)--(F) node at (1.1,-2.95) {$x_{1,1}$};
    
    \node (E2) at (4.5,0) {$\bullet$};
    \node[right=of E2] (F2) {$\bullet$};
    \node[below=of F2] (A2) {$\bullet$};
    \node[below=of E2] (B2) {$\bullet$};
    \draw[->,ultra thick] (E2)--(A2) node at (5.6,-2.95) {$x_{1,2}$};;
    
    \node (E3) at (9,0) {$\bullet$};
    \node[right=of E3] (F3) {$\bullet$};
    \node[below=of F3] (A3) {$\bullet$};
    \node[below=of E3] (B3) {$\bullet$};
    \draw[->,ultra thick] (B3)--(F3) node at (10.1,-2.95) {$x_{2,1}$};;
    
    \node (E4) at (13.5,0) {$\bullet$};
    \node[right=of E4] (F4) {$\bullet$};
    \node[below=of F4] (A4) {$\bullet$};
    \node[below=of E4] (B4) {$\bullet$};
    \draw[->,ultra thick] (B4)--(A4) node at (14.6,-2.95) {$x_{2,2}$};;
    
    \node (E5) at (0,-5) {$\bullet$};
    \node[right=of E5] (F5) {$\bullet$};
    \node[below=of F5] (A5) {$\bullet$};
    \node[below=of E5] (B5) {$\bullet$};
    \draw[->,ultra thick] (E5)--(A5) node at (3.1,-6.2) {\Large $\ast$};;
    
    \node (E6) at (4,-5) {$\bullet$};
    \node[right=of E6] (F6) {$\bullet$};
    \node[below=of F6] (A6) {$\bullet$};
    \node[below=of E6] (B6) {$\bullet$};
    \draw[->,ultra thick] (B6)--(F6);; 
    
    \node (E7) at (8,-5) {$\bullet$};
    \node[right=of E7] (F7) {$\bullet$};
    \node[below=of F7] (A7) {$\bullet$};
    \node[below=of E7] (B7) {$\bullet$};
    \draw[->,ultra thick] (E7)--(F7) node at (7.1,-6.2) {\Large $=$};;
    
    \node (E8) at (0,-9) {$\bullet$};
    \node[right=of E8] (F8) {$\bullet$};
    \node[below=of F8] (A8) {$\bullet$};
    \node[below=of E8] (B8) {$\bullet$};
    \draw[->,ultra thick] (E8)--(F8) node at (3.1,-10.2) {\Large $\ast$};;
    
    \node (E9) at (4,-9) {$\bullet$};
    \node[right=of E9] (F9) {$\bullet$};
    \node[below=of F9] (A9) {$\bullet$};
    \node[below=of E9] (B9) {$\bullet$};
    \draw[->,ultra thick] (B9)--(F9)  node at (7.1,-10.3) {\Large $=$};;
    \node (E10) at (8.1,-10.2) {\Large $0$};
    
    \draw [ultra thick, decorate,decoration={brace,amplitude=10pt,mirror},xshift=130.4pt,yshift=-0.4pt](6.5,-11.5) -- (6.5,-4.5) node[black,midway,xshift=2.5cm] {\footnotesize Composition rules on $\cS^2_1$};
\end{tikzpicture}

	\caption{The elements of $\cS^2_1$ and their composition rule, obtained by tracing arrows. If the arrows cannot be traced in a continuous manner the resulting element is 0. }
\label{s21fig}
\end{figure}
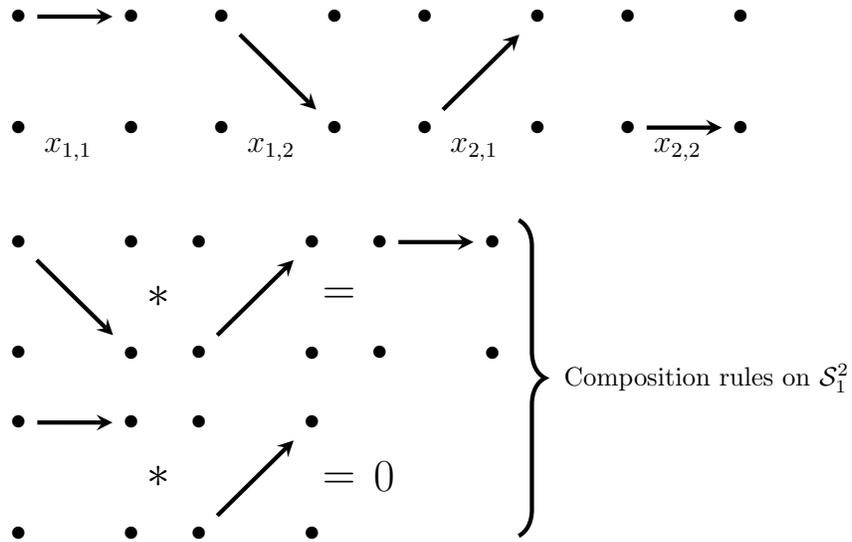

The simplest case is $\cS^2_1$, whose diagrammatics are shown in Fig. \ref{s21fig}. The partial symmetry elements of $\cS^2_1$ are denoted by $x_{a,b}$ with $a,b\in\{1,2\}$, and obey the following composition rule
\begin{equation}
\label{comprule}
 x_{a,b}\ast x_{c,d} = \delta_{bc}\,x_{a,d}.
\end{equation}
The indices $a$ and $b$ can be thought of, respectively, as the domain and range of the partial symmetry operation. The product between these elements is null when the range of the first element is different from the domain of the second element it is being composed with. Note that this product is non-commutative. 

Another example is $\cS^3_1$, which consists of nine elements $x_{a,b}$ with $a,b\in\{1,2,3\}$, as shown in Fig. \ref{s41}, with the same composition rule \eqref{comprule}.
\begin{figure}[ht!]
\centering
\begin{tikzpicture}[scale=.80]
    \node (A) at (0,0) {$\bullet$};
    \node (B) at (1.5,0) {$\bullet$};
    \node (C) at (1.5,0.5) {$\bullet$};
    \node (D) at (0,0.5) {$\bullet$};
    \node (E) at (1.5,1.0) {$\bullet$};
    \node (F) at (0,1.0) {$\bullet$};
    \draw[->,ultra thick] (F)--(E) node at (0.75,-0.5) {$x_{1,1}$};

    \node (A0) at (3,0) {$\bullet$};
    \node (B0) at (4.5,0) {$\bullet$};
    \node (C0) at (4.5,0.5) {$\bullet$};
    \node (D0) at (3,0.5) {$\bullet$};
    \node (E0) at (4.5,1.0) {$\bullet$};
    \node (F0) at (3,1.0) {$\bullet$};
    \draw[->,ultra thick] (F0)--(C0) node at (3.75,-0.5) {$x_{1,2}$};
    
    \node (A1) at (6,0) {$\bullet$};
    \node (B1) at (7.5,0) {$\bullet$};
    \node (C1) at (7.5,0.5) {$\bullet$};
    \node (D1) at (6,0.5) {$\bullet$};
    \node (E1) at (7.5,1.0) {$\bullet$};
    \node (F1) at (6,1.0) {$\bullet$};
    \draw[->,ultra thick] (F1)--(B1) node at (6.75,-0.5) {$x_{1,3}$};
    
    \node (A2) at (0,-1.5) {$\bullet$};
    \node (B2) at (1.5,-1.5) {$\bullet$};
    \node (C2) at (1.5,-2.0) {$\bullet$};
    \node (D2) at (0,-2.0) {$\bullet$};
    \node (E2) at (1.5,-2.5) {$\bullet$};
    \node (F2) at (0,-2.5) {$\bullet$};
    \draw[->,ultra thick] (D2)--(B2) node at (0.75,-3.0) {$x_{2,1}$};

    \node (A3) at (3,-1.5) {$\bullet$};
    \node (B3) at (4.5,-1.5) {$\bullet$};
    \node (C3) at (4.5,-2) {$\bullet$};
    \node (D3) at (3,-2) {$\bullet$};
    \node (E3) at (4.5,-2.5) {$\bullet$};
    \node (F3) at (3,-2.5) {$\bullet$};
    \draw[->,ultra thick] (D3)--(C3) node at (3.75,-3) {$x_{2,2}$};
    
    \node (A4) at (6,-1.5) {$\bullet$};
    \node (B4) at (7.5,-1.5) {$\bullet$};
    \node (C4) at (7.5,-2) {$\bullet$};
    \node (D4) at (6,-2) {$\bullet$};
    \node (E4) at (7.5,-2.5) {$\bullet$};
    \node (F4) at (6,-2.5) {$\bullet$};
    \draw[->,ultra thick] (D4)--(E4) node at (6.75,-3) {$x_{2,3}$};
    
    \node (A5) at (0,-4) {$\bullet$};
    \node (B5) at (1.5,-4) {$\bullet$};
    \node (C5) at (1.5,-4.5) {$\bullet$};
    \node (D5) at (0,-4.5) {$\bullet$};
    \node (E5) at (1.5,-5.0) {$\bullet$};
    \node (F5) at (0,-5.0) {$\bullet$};
    \draw[->,ultra thick] (F5)--(B5) node at (0.75,-5.5) {$x_{3,1}$};

    \node (A6) at (3,-4) {$\bullet$};
    \node (B6) at (4.5,-4) {$\bullet$};
    \node (C6) at (4.5,-4.5) {$\bullet$};
    \node (D6) at (3,-4.5) {$\bullet$};
    \node (E6) at (4.5,-5.0) {$\bullet$};
    \node (F6) at (3,-5.0) {$\bullet$};
    \draw[->,ultra thick] (F6)--(C6) node at (3.75,-5.5) {$x_{3,2}$};
    
    \node (A7) at (6,-4) {$\bullet$};
    \node (B7) at (7.5,-4) {$\bullet$};
    \node (C7) at (7.5,-4.5) {$\bullet$};
    \node (D7) at (6,-4.5) {$\bullet$};
    \node (E7) at (7.5,-5.0) {$\bullet$};
    \node (F7) at (6,-5.0) {$\bullet$};
    \draw[->,ultra thick] (F7)--(E7) node at (6.75,-5.5) {$x_{3,3}$};

  \end{tikzpicture}

\caption{The elements of $\cS^3_1$.}
\label{s41}
\end{figure}
This construction naturally generalizes to an $\cS^n_p$ with arbitrary $n$ and $p$, with $p$ denoting the number of arrows in the elements. In this paper we restrict our attention to the case of $p=1$ and generic $n$. We will see that $n$ controls the dimensionality of the local Hilbert space $d$. 

To explicitly see how to realize supersymmetry in terms of SISs let us start with $\cS^2_1$, building the supercharges $q$ and $q^\dag$ as 
\begin{equation}
\label{s21}
q = x_{1,2},\qquad q^\dag=x_{2,1},
\end{equation}
which are automatically nilpotent because of \eqref{comprule}. The Hamiltonian is $h = x_{1,1} + x_{2,2}$, with $b=x_{1,1}$ and $f=x_{2,2}$. Let us represent $\cS^2_1$ on a two-dimensional qubit space $\mathbb{C}^2$, with  basis spanned by $\{\ket{0}, \ket{1}\}$:
\begin{equation}
q = \left(\begin{array}{cc} 0 & 1 \\ 0 & 0 \end{array}\right),\qquad  q^\dag = \left(\begin{array}{cc} 0 & 0 \\ 1 & 0 \end{array}\right).
\end{equation}
In this particular case the resulting Hamiltonian is trivially the identity. The Witten operator can be constructed using \eqref{witten2} and it is easily seen to satisfy \eqref{witten3}. 

A non-trivial supersymmetric Hamiltonian is obtained if one starts instead from $\cS^3_1$ and considers
\begin{equation}\label{s31}
q = \frac{1}{\sqrt{2}}\left[x_{1,2} + x_{1,3}\right],\qquad q^\dag = \frac{1}{\sqrt{2}}\left[x_{2,1} + x_{3,1}\right].
\end{equation}
This results in a projector Hamiltonian as in \eqref{sissusy}
\begin{equation}\label{h31}
h = b + f = x_{1,1} + \frac{1}{2}\left[x_{2,2} + x_{2,3} + x_{3,2} + x_{3,3}\right].
\end{equation}
Representing $\cS^3_1$ on the three-dimensional qutrit space spanned by $\{\ket{0}, \ket{1}, \ket{2}\}$, one sees that $h$ is no longer the identity, but the sum of projectors to the two different sectors of the three-dimensional space: the one-dimensional bosonic sector spanned by $\ket{0}$ and the two-dimensional fermionic sector spanned by $\{\ket{1}, \ket{2}\}$. 

The systems with $\cS^d_1$ realizations can also be interpreted as non-supersymmetric spin-$\frac{d-1}{2}$-chains by regarding $x_{a,b}$ with $a>b$ ($a<b$) as spin-raising (lowering) 
operators. Then, $x_{a,a}$ stands for a projection operator to a state of spin $(a-\frac{d+1}{2})$.   

\section{The SLOCC classification of multi-qubit states  }
\label{SLOCCclasses}

Before proceeding to the construction of $R$-matrices using supersymmetry, we take a look at the different types of entangled states in a multi-qubit space. This is going to be useful later to clarify the role of supersymmetry in the classification of such states and as a guide for finding the relevant $R$-matrices that generate them.

A state $\ket{\phi}\in{\cal H}_\textrm{total}$ can be converted to another state $\ket{\psi}\in{\cal H}_\textrm{total}$ through {\it Stochastic Local Operations and Classical Communication} (SLOCC) when there exists an $N$-party protocol that allows any number of local quantum operations, ${\cal O}_i:\mathcal{H}_i\rightarrow\mathcal{H}_i$, along with classical communication among the $N$ parties. These local operations can also be projective measurements or unitary operators in extended systems. In this case we denote $\ket{\phi}\preceq\ket{\psi}$. This is a preorder relation and it induces an equivalence relation among states \cite{coecke}.

With this definition two states $\ket{\psi}$ and $\ket{\phi}$ are SLOCC-equivalent if and only if there exists an {\it invertible local operator} (ILO) such that 
\begin{equation}
\ket{\psi} = \left(L_1\otimes\cdots\otimes L_N\right)\ket{\phi},
\end{equation}
with $L_i:\mathcal{H}_i\rightarrow\mathcal{H}_i$~\cite{BPRS}. 
In this case we denote $\ket{\psi}\sim\ket{\phi}$, by which we classify multi-qubit states into different equivalence classes. 
Measurements connect different SLOCC classes as they are carried out through non-invertible operators and in general they reduce the amount of entanglement in the state. 

\paragraph{Two qubits}

There are two SLOCC classes in a system of two qubits: a class of the Bell states and a class of product states. Calling the qubits $A$ and $B$, the two classes are denoted by $AB$, for entangled qubits, and by $A-B$, for unentangled ones. 

Let us verify the statement above. The four Bell states are given by 
\begin{eqnarray}
\label{bell_states}
\ket{\psi_1}&=&\frac{1}{\sqrt{2}}\left[\ket{0,0}-\ket{1,1}\right], \qquad
\ket{\psi_2}= \frac{1}{\sqrt{2}}\left[\ket{0,1}-\ket{1,0}\right], \cr
\ket{\psi_3}&=& \frac{1}{\sqrt{2}}\left[\ket{0,1}+\ket{1,0}\right], \qquad
\ket{\psi_4}= \frac{1}{\sqrt{2}}\left[\ket{0,0}+\ket{1,1}\right].  	
\end{eqnarray}
Clearly, they are all SLOCC-equivalent as, for example, $\sigma^x_B\ket{\psi_1}  = \ket{\psi_2}$, which is an ILO ($\sigma^x_B$ is the Pauli $\sigma^x$ acting on $B$). Each one of these states can be converted into the others by similar ILOs. Moreover, any generic entangled two-qubit state is SLOCC-equivalent to a Bell state. For example, 
\begin{equation}
\sqrt{2}\left(\begin{array}{cc} k_1 & k_3 \\ k_2 & k_4\end{array}\right)_B \ket{\psi_4} = k_1\ket{0,0} + k_2\ket{0,1} + k_3\ket{1,0} + k_4\ket{1,1},
\end{equation}
which is the most arbitrary entangled two-qubit state when $k_1k_4 \neq k_2k_3$. 

On the other hand, the product basis $\{\ket{0,0}, \ket{0,1}, \ket{1,0}, \ket{1,1}\}$ of the $A-B$ SLOCC class is obtained from the Bell basis by a measurement. For example,
\begin{equation}
\ket{0,0}  = \frac{1}{\sqrt{2}}(1+\sigma^z_A)\ket{\psi_1}.
\end{equation}
Finally, the states in the product basis are all SLOCC-equivalent to each other. 

\paragraph{Three qubits}

There are now six different SLOCC classes \cite{Cirac}. Two of them are tripartite entangled states: the GHZ states and the W-states. There are three kinds of bipartite entangled states: $AB-C$, $A-BC$ and $AC-B$.  The sixth class are the unentangled product states, $A-B-C$. 

The GHZ states are built out of the product basis $\{\ket{\phi_j},\ket{\bar{\phi}_j}; 1\leq j\leq 4\}$ as
\begin{equation}
\label{ghz_states}
\ket{\psi^{\pm}_j} = \frac{1}{\sqrt{2}}\left[\ket{\phi_j}\pm\ket{\bar{\phi}_j}\right],\qquad 1\leq j\leq 4,
\end{equation}
where $\ket{\bar{\phi}_j}$ is obtained from  $\ket{\phi_j}$ by interchanging 0 and 1 on every site. For example, for $\ket{\phi_1}=\ket{000}$, $\ket{\phi_2}=\ket{100}$, we have $\ket{\bar{\phi}_1}=\ket{111}$, $\ket{\bar{\phi}_2}=\ket{011}$. This includes the standard state $\ket{\psi_1^+} = \left[\ket{000}+\ket{111}\right]/\sqrt{2}$. 

The other inequivalent tripartite class is the W-state class comprising
\begin{eqnarray} 
\ket{w_1} & = & \frac{1}{\sqrt{3}}\left[\ket{100} + \ket{010} + \ket{001}\right], \qquad
\ket{w_2}  =  \frac{1}{\sqrt{3}}\left[\ket{101} + \ket{011} + \ket{000}\right],\cr
\ket{w_3} & = & \frac{1}{\sqrt{3}}\left[\ket{110} + \ket{000} - \ket{011}\right], \qquad
\ket{w_4}  =  \frac{1}{\sqrt{3}}\left[\ket{000} - \ket{110} - \ket{101}\right], \cr
\ket{w_5} & = & \frac{1}{\sqrt{3}}\left[\ket{111} + \ket{001} - \ket{010}\right], \qquad
\ket{w_6}  =  \frac{1}{\sqrt{3}}\left[\ket{001} - \ket{111} - \ket{100}\right], \cr
\ket{w_7} & = & \frac{1}{\sqrt{3}}\left[\ket{010} - \ket{100} + \ket{111}\right], \qquad
\ket{w_8}  =  \frac{1}{\sqrt{3}}\left[\ket{011} - \ket{101} + \ket{110}\right]. 
\label{w1}\label{w8}\label{w2}
\end{eqnarray}
An arbitrary superposition in each state is SLOCC-equivalent to the standard form. For example, the state $\alpha\ket{100}+\beta\ket{010}+\gamma\ket{001}$ 
is SLOCC-equivalent to the first standard W-state in \eqref{w1} since
\begin{equation}
\sqrt{3} \left(\begin{array}{cc}1&0\\0&\alpha\end{array}\right)_A\left(\begin{array}{cc}1&0\\0&\beta\end{array}\right)_B\left(\begin{array}{cc}1&0\\0&\gamma\end{array}\right)_C~\ket{w_1}= \alpha\ket{100}+\beta\ket{010}+\gamma\ket{001}.
\end{equation}
We can obtain the ILOs for the other states in a similar manner by inspection. For a geometric way of obtaining these states see \cite{geomqstates}. It is shown in \cite{Cirac} that the W-state class and the GHZ state class are not SLOCC-equivalent.

The bipartite entangled class is formed by states where two of the qubits are in the Bell state class. For example, a state in $A-BC$ is given by $\left[\ket{000}+\ket{011}\right]/\sqrt{2}$. One can write the four Bell states in each of the three classes $A-BC$, $AB-C$ and $AC-B$. 

\paragraph{Multi-qubits }

For four or more qubits the SLOCC classification gets much harder as there is an infinite number of classes \cite{Cirac}. However, the GHZ state class and the W-state class have a natural generalization to these cases. We will just write down the states in these two classes so that we can identify the states obtained from the $R$-matrices later. 

 The standard product basis for $\mathcal{H}^{\otimes N}$ is denoted by $\{\ket{\phi_j},\ket{\bar{\phi}_j}; 1\leq j\leq 2^{N-1}\}$.
The GHZ states are given by 
\begin{equation}
\label{GHZ states}
\ket{\psi^{\pm}_j} = \frac{1}{\sqrt{2}}\left[\ket{\phi_j}\pm\ket{\bar{\phi}_j}\right],\qquad 1\leq j\leq 2^{N-1},
\end{equation}
with $\ket{\bar{\phi}}$ obtained from $\ket{\phi}$ by flipping 0 and 1 on every site.

The W-state class, on the other hand, can be constructed by acting with the following unitary matrix on the standard product basis:
\begin{equation}\label{majorana}
U = \frac{1}{\sqrt{N}}\sum_{j=1}^N~\chi_j,
\end{equation}
where $\chi_k  = \left(\prod_{j=1}^{k-1}\sigma^z_j\right) \sigma^x_k$ satisfies $\chi_k^2=1$ and $\chi_{k}\chi_{l}=-\chi_{l}\chi_{k}$ ($k\neq l$). It is easy to see that $U^2=1$ and $U^\dag=U$. Acting on $\ket{00\cdots0}$ produces the $N$-qubit W-state. 

\section{$R$-matrices from supersymmetry}
\label{susyR}

Having laid out this groundwork, we are now ready to use supersymmetry, locally realized via SISs, to construct solutions of the spectral parameter-dependent $(d,m,l)$-gYBE. We start with finding non-unitary $R$-matrices\footnote{Non-unitary $R$-matrices are discussed in \cite{MMMMM} in the context of topological quantum computation.} and then move on to the unitary ones. Moreover, we shall see that previously known solutions, like the  solution by Rowell, Wang and collaborators \cite{r1,r2}, can also be obtained through this method. 

\subsection{Non-unitary solutions}

We start by considering the  case of two ($m=2$) and three ($m=3$) qubits or qudits, and then generalize to an arbitrary number of them. The dimensionality $d$ of the local Hilbert space is going to be selected by the particular choice of SISs, with $\cS^n_1$ fixing $d=n$.

\subsubsection*{Two qubits ($m=2$, $l=1$)} 

The supersymmetry algebra allows to easily construct $R$-matrices of the form \eqref{Rqi}. There are many ways of doing this. All the following combinations of supercharges and Witten operators
\begin{equation}
\label{Schoices}
Q_i = w_iq_{i+1}, \quad Q_i = q_iw_{i+1}, \quad
Q_i = q_iq_{i+1},\quad Q_i=q^\dag_iq^\dag_{i+1} , \quad
Q_i = q_iq^\dag_{i+1},\quad Q_i = q^\dag_iq_{i+1}
\end{equation}
satisfy $Q_i^2=0$ and $Q_iQ_{i+1}Q_i = Q_{i+1}Q_iQ_{i+1} = 0$. 

It is now a matter of choosing a particular representation of these operators in terms of SISs. Starting with the simple case of $\cS^2_1$, one can take the supercharges to be given by $q_i=\left(x_{1,2}\right)_i$ and $q_i^\dag=\left(x_{2,1}\right)_i$. As mentioned earlier the ``2" in $\cS^2_1$ fixes the dimensionality of the local Hilbert space. The indices 1 and 2 in the SIS variables correspond to the qubits $\ket{0}$ and $\ket{1}$, respectively. The $R$-matrix in \eqref{Rqi} with $Q_i=q_iq_{i+1}$ gives a state $\ket{11} + c u\ket{00}$ upon acting on $\ket{11}$, while leaving the other product states invariant. This state coincides with $\ket{\psi_1}$ or $\ket{\psi_4}$ in \eqref{bell_states}, up to weights of the superpositions. As seen above, it is SLOCC-equivalent to the standard form of the Bell states (\ref{bell_states}). In a similar fashion, using the $R$-matrix built out of $Q_i =q_i^\dag q_{i+1}$ one gets the other Bell states containing $\ket{01}$ and $\ket{10}$. The $R$-matrices from $Q_i=q_iw_{i+1}$ and $Q_i=w_iq_{i+1}$ just give product states, thus exhausting the two SLOCC classes in the two-qubit space. 

More explicitly, the two-qubit non-unitary $R$-matrix built out of $Q_i=q_iq_{i+1}$ in (\ref{Schoices}) is given by 
\begin{equation}
R_i(u) = \left(\begin{array}{cccc} 1 & 0 & 0 & c u \\ 0 & 1 & 0 & 0\\ 0 & 0 & 1 & 0\\ 0 & 0 & 0 & 1 \end{array}\right).
\end{equation} 
This reproduces a result in \cite{vieira} derived from a different approach (equation (A.14) there times the permutation matrix coincides with our result).

Moving on to $\cS^3_1$, one can take $q_i = \left[\left(x_{1,2}\right)_i + \left(x_{1,3}\right)_i\right]/\sqrt{2}$ and $q_i^\dag = \left[\left(x_{2,1}\right)_i + \left(x_{3,1}\right)_i\right]/\sqrt{2}$, as done in \eqref{s31}. This is now a local Hilbert space of dimension 3, {\it i.e.} a qutrit space spanned by $\{\ket{0}, \ket{1}, \ket{2}\}$. Nevertheless, we still produce Bell-like states as the chosen supercharges grade the Hilbert space into a one-dimensional bosonic part spanned by $\ket{0}$ and a two-dimensional fermionic part spanned by $\ket{1}$ and $\ket{2}$. The local supercharge $q_i$ acts only on $\{\ket{1}, \ket{2}\}$, converting them into the lone boson $\ket{0}$, while the adjoint $q_i^\dag$ does the reverse. Hence the $R$-matrix built out of $Q_i = q_i^\dag q_{i+1}^\dag$ produces the Bell-like state of qutrits $\ket{00} + c u\ket{\tilde{1}\tilde{1}}$, with $\ket{\tilde{1}} = \left[\ket{1}+\ket{2}\right]/\sqrt{2}$. As in the $\cS^2_1$ realization, one can obtain the other Bell-like state $\ket{0\tilde{1}} + c u\ket{\tilde{1}0}$ using $Q_i=q_i^\dag q_{i+1}$, whereas the product states are obtained  using choices in \eqref{Schoices} containing the Witten operators. This exhausts all Bell-like states in the two-qutrit system. 
 
This easily generalizes to the qudit case by using an $\cS^d_1$ realization of the supercharges, allowing to construct Bell-like states in the two-qudit space. 

\subsubsection*{Three qubits ($m=3,~l=1$)} 

We look again at solutions of the form \eqref{Rqi} and separate the solutions according to the SLOCC class of states they produce. All the operators $Q_i$ we write below can be checked to satisfy $Q_i^2=0$ and $Q_iQ_{i+1}Q_i=Q_{i+1}Q_iQ_{i+1}=0$. 

The following choices
\begin{eqnarray}
Q_i &=& w_iw_{i+1}q_{i+2}, \qquad Q_i=w_iw_{i+1}q_{i+2}^\dag, \qquad
Q_i = w_iq_{i+1}w_{i+2}, \cr 
Q_i & =& w_iq_{i+1}^\dag w_{i+2}, \qquad
Q_i = q_iw_{i+1}w_{i+2}, \qquad Q_i=q_i^\dag w_{i+1}w_{i+2}. 
\end{eqnarray}
can be easily seen to produce product states in a three-qudit space, by using an explicit $\cS^d_1$ realization of the supercharges. In the language of SLOCC classes these are states of the form $A-B-C$.

Partially entangled states of the $AB-C$, $A-BC$ and $AC-B$ classes, respectively, are obtained by using 
\begin{eqnarray}
\label{QPE}
Q_i  &=&  q_iq_{i+1} w_{i+2},\qquad Q_i  =  q_i^\dag q_{i+1}^\dag w_{i+2}, \qquad 
Q_i  =  q_i^\dag q_{i+1}w_{i+2},\qquad Q_i  =  q_iq_{i+1}^\dag w_{i+2};\cr
Q_i &=&  w_iq_{i+1} q_{i+2},\qquad Q_i  =  w_i q_{i+1}^\dag q_{i+2}^\dag, \qquad
Q_i  =  w_i q_{i+1}q_{i+2}^\dag,\qquad Q_i  =  w_iq_{i+1}^\dag q_{i+2}; \cr
Q_i & = & q_iw_{i+1} q_{i+2},\qquad Q_i  =  q_i^\dag w_{i+1} q_{i+2}^\dag, \qquad
Q_i  =  q_i w_{i+1}q_{i+2}^\dag,\qquad Q_i  =  q_i^\dag w_{i+1} q_{i+2}. 
\end{eqnarray}
As in the earlier cases, an $\cS^d_1$ realization of the supercharges produces the partially entangled SLOCC classes of two parties in the three-qudit space. Explicitly, the non-unitary $R$-matrix producing a partially entangled state built out of $Q_i = q_iq_{i+1}w_{i+2}$ in (\ref{QPE}) is given by
\begin{equation}
R_i(u) = \left(\begin{array}{cccccccc} 1 & 0 & 0 & 0 & 0 & 0 & -c u & 0 \\ 0 & 1 & 0 & 0 & 0 & 0 & 0 & c u \\ 0 & 0 & 1 & 0 & 0 & 0 & 0 & 0 \\
0 & 0 & 0 & 1 & 0 & 0 & 0 & 0 \\ 0 & 0 & 0 & 0 & 1 & 0 & 0 & 0 \\ 0 & 0 & 0 & 0 & 0 & 1 & 0 & 0 \\ 0 & 0 & 0 & 0 & 0 & 0 & 1 & 0 \\ 0 & 0 & 0 & 0 & 0 & 0 & 0 & 1 \end{array}\right).
\end{equation} 

The eight GHZ states \eqref{ghz_states} are produced by
\begin{eqnarray}
\label{QGHZ}
Q_i & = & q_iq_{i+1} q_{i+2},\qquad  Q_i=q_iq_{i+1} q_{i+2}^\dag,\qquad
Q_i  =  q_iq_{i+1}^\dag q_{i+2},\qquad Q_i=q_i^\dag q_{i+1} q_{i+2},\cr
Q_i & = & q_i^\dag q_{i+1}^\dag q_{i+2},\qquad Q_i=q_iq_{i+1}^\dag q_{i+2}^\dag,\qquad
Q_i  =  q_i^\dag q_{i+1} q_{i+2}^\dag,\qquad  Q_i=q_i^\dag q_{i+1}^\dag q_{i+2}^\dag
\end{eqnarray}
with the $\cS^2_1$ realization, up to weights of the superpositions. The non-unitary $R$-matrix generating the state built out of $Q_i=q_iq_{i+1}q_{i+2}^\dag$ in the equation above is
\begin{equation}
R_i(u) = \left(\begin{array}{cccccccc} 1 & 0 & 0 & 0 & 0 & 0 & 0 & 0 \\ 0 & 1 & 0 & 0 & 0 & 0 & c u & 0 \\ 0 & 0 & 1 & 0 & 0 & 0 & 0 & 0 \\
0 & 0 & 0 & 1 & 0 & 0 & 0 & 0 \\ 0 & 0 & 0 & 0 & 1 & 0 & 0 & 0 \\ 0 & 0 & 0 & 0 & 0 & 1 & 0 & 0 \\ 0 & 0 & 0 & 0 & 0 & 0 & 1 & 0 \\ 0 & 0 & 0 & 0 & 0 & 0 & 0 & 1 \end{array}\right).
\end{equation}

The W-state $\ket{w_1}$ in \eqref{w1} is constructed from
\begin{equation}
Q_i = b_iq_{i+1}^\dag q_{i+2} + q_i^\dag b_{i+1}q_{i+2},
\label{Qw1}
\end{equation}
while the remaining W-states $\ket{w_2}$ to $\ket{w_8}$ are built using
\begin{eqnarray}
Q_i & = &  q_i^\dag b_{i+1} q_{i+2}^\dag + b_i q_{i+1}^\dag q_{i+2}^\dag, \qquad
Q_i  =   q_i^\dag q_{i+1}^\dag b_{i+2} + b_i q_{i+1}^\dag q_{i+2}^\dag, \cr
Q_i & = &  q_i^\dag q_{i+1}^\dag b_{i+2} + q_i^\dag b_{i+1} q_{i+2}^\dag, \qquad
Q_i  =   q_i q_{i+1} f_{i+2} + q_i f_{i+1} q_{i+2}, \cr
Q_i & = &  q_i q_{i+1} f_{i+2} + f_i q_{i+1} q_{i+2}, \qquad
Q_i  =   q_i f_{i+1} q_{i+2} + f_i q_{i+1} q_{i+2}, \cr
Q_i & = &  q_i^\dag q_{i+1} f_{i+2} + q_i^\dag f_{i+1} q_{i+2},
\end{eqnarray}
respectively. 
The explicit form of the $R$-matrix from (\ref{Qw1}) is 
\begin{equation}
R_i(u) = \left(\begin{array}{cccccccc} 1 & 0 & 0 & 0 & 0 & 0 & 0 & 0 \\ 0 & 1 & 0 & 0 & 0 & 0 & 0 & 0 \\ 0 & cu & 1 & 0 & 0 & 0 & 0 & 0 \\ 0 & 0 & 0 & 1 & 0 & 0 & 0 & 0 \\ 
0 & cu & 0 & 0 & 1 & 0 & 0 & 0 \\ 0 & 0 & 0 & 0 & 0 & 1 & 0 & 0 \\ 0 & 0 & 0 & 0 & 0 & 0 & 1 & 0 \\ 0 & 0 & 0 & 0 & 0 & 0 & 0 & 1 \end{array}\right). 
\end{equation}
The $R$-matrices built from these operators produce the W-states with coefficients depending on the spectral parameter $u$. However, such states are in the same SLOCC class as the W-states in \eqref{w1}.  These considerations exhaust the SLOCC classes in the three-qubit sector. 

\subsubsection*{Multi-qubits ($m$, $l=1$)} 

The complete SLOCC classification for the multi-qubit case is unknown. However, some of the states from the three-qubit sectors are easily generalized. These include product states, the partially entangled states, the GHZ states and the W-states. We will write down the $R$-matrices that produce just these states. Again, it is easy to verify that $Q_i^2=0$ and  $Q_iQ_{i+1}Q_i=Q_{i+1}Q_iQ_{i+1}=0$ for each of the choices below.  

Product states are generated by the following $Q_i$ operators to be inserted in \eqref{Rqi}
\begin{eqnarray}
\label{mproduct_states}
Q_i^r  =  \left(\prod_{j=0}^{r-1}~w_{i+j}\right)q_{i+r}\left(\prod_{j=r+1}^{m-1}~w_{i+j}\right), \qquad 
Q_i^r  =  \left(\prod_{j=0}^{r-1}~w_{i+j}\right)q_{i+r}^\dag\left(\prod_{j=r+1}^{m-1}~w_{i+j}\right), 
\end{eqnarray}
for $r=0, 1, \cdots, m-1$, giving $2m$ different choices.

The $Q_i$ operators partially entangling $r$ qubits of the $m$-qubit system into an $A_1\cdots A_r-A_{r+1}-\cdots- A_m$ SLOCC class are given by
\begin{equation}\label{qir}
Q_i^{r;~(\alpha_1,\cdots, \alpha_r)} = \left(\prod_{j=0}^{r-1}~q_{i+j}^{\alpha_{j+1}}\right)\left(\prod_{j=r}^{m-1}~w_{i+j}\right),
\end{equation}
with each $\alpha_j\in \{\textrm{nothing}, \dag\}$, giving $2^r$ choices. Note that $r$ can take values in $\{2,\ldots, m-1\}$. The $R$-matrices from these $Q_i$ give the $r$-qubit GHZ states embedded in an $m$-qubit system. By permuting the $r$ supercharges in the $Q_i$ in \eqref{qir}, one obtains the other $r$-qubit partially entangled sectors in the $m$-qubit space. There are a total of $\left(\begin{array}{c}m \\ r\end{array}\right)$ such choices, corresponding to inequivalent SLOCC classes.

The $2^m$ $m$-qubit GHZ states are obtained from the $R$-matrices built out of the following
\begin{equation}
Q_i^{\alpha_1,\cdots, \alpha_m} = \prod_{j=0}^{m-1}~q_{i+j}^{\alpha_{j+1}},
\end{equation}
with each $\alpha_j\in\{\textrm{nothing}, \dag\}$. 

The $m$-qubit W-states are generated using a unitary operator as in \eqref{majorana}. We present the construction of just one of these standard states, namely $\sum_{r=1}^m~\ket{0\cdots 0 1_r0\cdots 0}$, which can be obtained from
\begin{equation}
\label{mWstates}
Q_i = \sum_{r=0}^{m-2}~\left(\prod_{j=0}^{r-1}~b_{i+j}\right)q_{i+r}^\dag\left(\prod_{j=r+1}^{m-2}~b_{i+j}\right)q_{i+m-1},
\end{equation}
where $b=qq^\dag$ is the projector to the bosonic sector. 

The cases above cover the most interesting multi-qubit states obtained from the $(2,m,1)$-gYBE solutions. We can extend these solutions to the $(d,m,1)$-gYBE case by choosing an $\cS^d_1$ realization for the supercharges. In fact, we can do even better by constructing solutions of the $(d,m,l)$-gYBE for arbitrary $l$ in \eqref{gYBEparameters}.

By increasing $l$ we are effectively changing the algebra of the $Q_i$ operators to 
\begin{equation}
\label{ql}
Q_iQ_{i+l}Q_i = Q_{i+l}Q_iQ_{i+l}.	
\end{equation}
When $l\geq m$ this is trivially satisfied as there is no overlap between $Q_i$ and $Q_{i+l}$. However, when $m>l$
the operators have a non-trivial overlap on $m-l$ sites. Nevertheless, \eqref{ql} is still satisfied by the $Q_i$ operators constructed to produce the multi-qubit states (\ref{mproduct_states})-(\ref{mWstates}), thus providing solutions for the $(d,m,l)$-gYBE.  

\subsection{Unitary solutions} 
\label{sec:unitary}

So far we have used non-hermitian $Q_i$ operators to build the $R$-matrices of the form \eqref{Rqi}.  The resulting $R$-matrices do not satisfy the unitarity condition
\begin{equation}
R^\dag_i(-u)R_i(u)=R_i(u)R^\dag_i(-u) = I.	
\end{equation}
A given supersymmetric system provides a number of hermitian operators constructed out of the supercharges $q$ and $q^\dag$, including the supersymmetric Hamiltonian and the projectors to the bosonic and fermionic sectors. We shall use these operators to build unitary $R$-matrices that generate the desired entangled states. As in the non-unitary case we will consider the $m=2$ and $m=3$ cases before generalizing to arbitrary $m$. 

\subsubsection*{The case $m=2$}

We start by constructing the $R$-matrices that produce the two classes in the two-qubit case: the Bell class $AB$ and the product states $A-B$. 

The Bell states are constructed as follows. Consider the supercharge (realized with $\cS^2_1$)
\begin{equation}\label{uq1}
Q_i = \frac{1}{\sqrt{\alpha^2+\beta^2}}\left[\alpha b_iq_{i+1} + \beta q_i^\dag f_{i+1}\right],
\end{equation}
where $\alpha, \beta\in \mathbb{R}$. It is clear that this grades $\mathbb{C}^2\otimes\mathbb{C}^2$ into a two-dimensional bosonic sector spanned by $\{\ket{00}, \ket{11}\}$ and a fermionic sector spanned by $\ket{01}$. 
$Q_i$ maps the fermionic sector to the bosonic sector, and $Q_i^\dag$ does the reverse. 
The state $\ket{10}$ is a zero-mode for this system as $Q_i\ket{10}=Q_i^\dag\ket{10}=0$. 
Consider also the projector to the bosonic sector
\begin{equation}
B_i=Q_iQ_i^\dag =\frac{1}{\alpha^2+\beta^2}\left[ \alpha^2 b_ib_{i+1} + \alpha\beta\left(q_iq_{i+1}+q_i^\dag q_{i+1}^\dag\right) + \beta^2 f_if_{i+1}\right],
\end{equation}
which maps the product states $\ket{00}$ and $\ket{11}$ to $\alpha\ket{00}+\beta\ket{11}$ up to the normalization, {\it i.e.} an SLOCC-equivalent state to the standard Bell state. We can construct an $R$-matrix of the form \eqref{Rcp} by making use of commuting projectors $B_i$ and $B_{i+l}$, obeying
\begin{equation}
B_iB_{i+l}B_i = B_{i+l}B_iB_{i+l}=B_iB_{i+l},\qquad l\geq 2.
\end{equation}
This gives a solution to the $(2,2,l)$-gYBE for all $l\geq 2$. 
The $R$-matrix leads to the entangled states
\begin{eqnarray}
\label{Rbell}
R_i(u)\ket{00} & = & \frac{1}{\alpha^2+\beta^2}\left[\left(\alpha^2e^{cu}+\beta^2\right)\ket{00} + \alpha\beta \left(e^{cu}-1\right)\ket{11}\right], \nonumber \\
R_i(u)\ket{11} & = & \frac{1}{\alpha^2+\beta^2}\left[\alpha\beta \left(e^{cu}-1\right)\ket{00} + \left(\alpha^2+\beta^2e^{cu}\right)\ket{11} \right],
\end{eqnarray}
where $c$ is a real constant. Note that in order to interpret $R_i(u)$ as a time-evolution operator, we should take $u$ as an imaginary time, namely $u=\textrm{i} t$, for unitary evolution. Then, the time evolution starting at $t=0$ drives the product states to entangled states. However, after integer multiples of the period $T=2\pi/c$, they come back to the product states. This is common to all the entangled states generated by $R_i(u)$ that we present below.  By using an $\cS^d_1$ realization of the supersymmetry, we also obtain a $(d,2,l)$-gYBE solution via the same operators. 

A two-qubit unitary $R$-matrix built out of the supercharge in (\ref{uq1}) is
\begin{equation}
R_i(u) = \left(\begin{array}{cccc} 1+\frac{\alpha^2}{\alpha^2+\beta^2}a(u) & 0 & 0 & \frac{\alpha\beta}{\alpha^2+\beta^2}a(u) \\ 0 & 1 & 0 & 0 \\ 0 & 0 & 1 & 0 \\ \frac{\alpha\beta}{\alpha^2+\beta^2}a(u) & 0 & 0 & 1 + \frac{\beta^2}{\alpha^2+\beta^2}a(u) \end{array}\right),
\end{equation} 
with $a(u) = \left(e^{cu}-1\right)$.  

We could have equally chosen another supercharge that grades the Hilbert space in a different way with the fermionic  sector now spanned by $\ket{10}$ and the bosonic sector remaining the same. The state $\ket{01}$ becomes the zero-mode. The supercharge that generates this system is
\begin{equation}
Q_i = \frac{1}{\sqrt{\alpha^2+\beta^2}}\left[\alpha q_ib_{i+1}+\beta f_iq_{i+1}^\dag\right],
\end{equation}
resulting in the same projector to the bosonic sector as for \eqref{uq1}. The $R$-matrix obtained this way only produces a weighted superposition of $\ket{00}$ and $\ket{11}$ when acting on $\ket{00}$ and $\ket{11}$. The other two product states $\ket{01}$ and $\ket{10}$ are left invariant. We can similarly produce the other Bell state $\ket{01}+\ket{10}$, as there is a canonical way of finding the right $Q_i$ to this scope. This supercharge must produce a grading of the Hilbert space such that  $\ket{01}$ and $\ket{10}$ belong to the bosonic sector and one of the other states, either $\ket{00}$ or $\ket{11}$, forms the fermionic sector. The remaining state is a zero-mode. If we select $\ket{11}$ to span the fermionic sector, the supercharge becomes
\begin{equation}\label{uq2}
Q_i =\frac{1}{\sqrt{\alpha^2+\beta^2}}\left[\alpha f_i q_{i+1} + \beta q_if_{i+1}\right],
\end{equation}
making the projector to the bosonic sector
\begin{equation}
B_i=Q_iQ_i^\dag = \frac{1}{\alpha^2+\beta^2}\left[\alpha^2 f_ib_{i+1} + \alpha\beta\left(q_i^\dag q_{i+1} + q_iq_{i+1}^\dag\right) + \beta^2 b_if_{i+1}\right].
\end{equation}
This commutes with $B_{i+2}$ and thus builds the $R$-matrix that produces the other Bell state 
\begin{eqnarray}
R_i(u)\ket{01} & = & \frac{1}{\alpha^2+\beta^2}\left[\left(\alpha^2+\beta^2e^{cu}\right)\ket{01} + \alpha\beta \left(e^{cu}-1\right)\ket{10}\right], \nonumber \\
R_i(u)\ket{10} & = & \frac{1}{\alpha^2+\beta^2}\left[\alpha\beta \left(e^{cu}-1\right)\ket{01} + \left(\alpha^2e^{cu}+\beta^2\right)\ket{10} \right].
\end{eqnarray}
We could have equally set the one dimensional fermionic sector to be spanned by the state $\ket{00}$, in which case the supercharge becomes 
\begin{equation}
Q_i =\frac{1}{\sqrt{\alpha^2+\beta^2}}\left[\alpha b_i q_{i+1}^\dag + \beta q_i^\dag b_{i+1}\right].
\end{equation}
This exhausts all the possibilities for producing the entangled SLOCC class of the two-qubit case.

The rationale for generating the Bell states above was to project onto a two-dimensional sector which could accommodate entangled qubits, which was achieved by using the bosonic projector. It is clear then that product states are going to be obtained by projecting onto the one-dimensional fermionic sector, using $F_i=Q_i^\dag Q_i$.
For example, the supercharge \eqref{uq1} gives $F_i = Q_i^\dag Q_i = b_if_{i+1}$, while the supercharge \eqref{uq2} gives $F_i = Q_i^\dag Q_i = f_if_{i+1}$. Clearly $F_i$ commutes with $F_{i+l}$, yielding $R$-matrices of the form \eqref{Rcp} that solve the $(d,2,l)$-gYBE for arbitrary $l$.

\subsubsection*{The case $m=3$}

As seen in Sec. \ref{SLOCCclasses}, the three-qubit case has six inequivalent SLOCC classes, for which we construct now the corresponding $R$-matrices. 

\paragraph{GHZ states} 
We start with GHZ states, by taking the supercharge 
\begin{equation}\label{uq3}
Q_i = \frac{1}{\sqrt{\alpha^2+\beta^2}}\left[\alpha b_iq_{i+1}q_{i+2} + \beta q_i^\dag f_{i+1}f_{i+2}\right].
\end{equation}
With the $\cS^2_1$ realization (\ref{s21}), the Hilbert space $\mathbb{C}^2\otimes\mathbb{C}^2\otimes\mathbb{C}^2$ gets graded into a two-dimensional bosonic sector spanned by $\{\ket{000}, \ket{111}\}$ and a one-dimensional fermionic sector spanned by $\ket{011}$. The unitary $R$-matrix is constructed from the projector to the bosonic sector given by
\begin{equation}\label{b3}
B_i = Q_iQ_i^\dag  = \frac{1}{\alpha^2+\beta^2}\left[\alpha^2 b_ib_{i+1}b_{i+2} + \alpha\beta\left(q_i q_{i+1}q_{i+2} + q_i^\dag q_{i+1}^\dag q_{i+2}^\dag\right) + \beta^2 f_if_{i+1}f_{i+2}\right].
\end{equation}
This projector commutes with $B_{i+l}$ implying 
\begin{equation}
B_iB_{i+l}B_i = B_{i+l}B_iB_{i+l}=B_iB_{i+l},\qquad l\geq 3,
\end{equation}
which leads to a unitary $R$-matrix that solves the $(2,3,l)$-gYBE for all $l\geq 3$. With an $\cS^d_1$ realization we obtain, as usual, solutions to the $(d,3,l)$-gYBE for all $l\geq 3$ via the same operators. The supercharge \eqref{uq3} is not the only choice that leads to the bosonic projector in \eqref{b3}. There are other equivalent supercharges that produce the same bosonic sector but a different one-dimensional fermionic sector, given by
\begin{eqnarray}
Q_i & = & \frac{1}{\sqrt{\alpha^2+\beta^2}}\left[\alpha b_ib_{i+1}q_{i+2} + \beta q_i^\dag q_{i+1}^\dag f_{i+2}\right], \cr
Q_i & = & \frac{1}{\sqrt{\alpha^2+\beta^2}}\left[\alpha b_iq_{i+1}b_{i+2} + \beta q_i^\dag f_{i+1} q_{i+2}^\dag\right], \cr
Q_i & = & \frac{1}{\sqrt{\alpha^2+\beta^2}}\left[\alpha q_ib_{i+1}b_{i+2} + \beta f_i q_{i+1}^\dag q_{i+2}^\dag\right], \cr
Q_i & = & \frac{1}{\sqrt{\alpha^2+\beta^2}}\left[\alpha q_ib_{i+1}q_{i+2} + \beta f_i q_{i+1}^\dag f_{i+2}\right], \cr
Q_i & = & \frac{1}{\sqrt{\alpha^2+\beta^2}}\left[\alpha q_iq_{i+1}b_{i+2} + \beta f_i f_{i+1} q_{i+2}^\dag\right], 
\end{eqnarray}
with $\ket{001}, \ket{010}, \ket{100}, \ket{101}, \ket{110}$ as the corresponding fermionic sectors. 
 
Note that this $R$-matrix only generates a state which is composed by the same product states as the standard GHZ state $\left[\ket{000}+\ket{111}\right]/\sqrt{2}$.  In order to generate another GHZ state in the same SLOCC class, such as $\left[\ket{001}+\ket{110}\right]/\sqrt{2}$, one can construct a supercharge that generates a two-dimensional bosonic sector spanned by the same product states as in the target entangled state, namely $\ket{001}$ and $\ket{110}$. The one-dimensional fermionic sector is spanned by the product state that gets converted to the entangled state by the supercharge $Q_i$. For example, the supercharge that converts the product states $\ket{000}$ and $\ket{111}$ into the desired GHZ state is given by
\begin{equation}
Q_i = \frac{1}{\sqrt{\alpha^2+\beta^2}}\left[\alpha b_ib_{i+1}q_{i+2}^\dag + \beta q_i^\dag q_{i+1}^\dag b_{i+2}\right].
\end{equation}
This generates the bosonic projector
\begin{equation}
B_i = \frac{1}{\alpha^2+\beta^2}\left[\alpha^2 b_ib_{i+1}f_{i+2} + \alpha\beta\left(q_i q_{i+1}q_{i+2}^\dag + q_i^\dag q_{i+1}^\dag q_{i+2}\right) + \beta^2 f_if_{i+1}b_{i+2}\right],
\end{equation}
which commutes with $B_{i+l}$ and builds a unitary $R$-matrix that solves the $(2,3,l)$-gYBE, for all $l\geq 3$, for the $\cS^2_1$ realization of the supercharge. 
Finally, we obtain 
\begin{eqnarray}
R_i(u)\ket{000} & = &  \frac{1}{\alpha^2+\beta^2}\left[\left(\alpha^2e^{cu}+\beta^2\right)\ket{000} + \alpha\beta \left(e^{cu}-1\right)\ket{111}\right], \nonumber \\
R_i(u)\ket{111} & = & \frac{1}{\alpha^2+\beta^2}\left[\alpha\beta \left(e^{cu}-1\right)\ket{000} + \left(\alpha^2+\beta^2e^{cu}\right)\ket{111} \right].
\end{eqnarray}
This $B_i$ also leads to a unitary $R$-matrix for the $(d,3,l)$-gYBE, for all $l\geq 3$, upon an $\cS^d_1$ realization of the supercharges. 
This method further elucidates the canonical way to construct the supercharge to produce the $R$-matrix that generates the desired entangled state. In a similar manner one can construct each of the other entangled GHZ states in the GHZ SLOCC class for the three-qubit system. 

\paragraph{Product states}
As in the two-qubit sector, we can construct the $R$-matrix that produces product states out of the projectors to the fermionic sector from the supercharges that produce GHZ-class states. For example, one such projector generated from the supercharge in \eqref{uq3} is $F_i= Q_i^\dag Q_i = b_if_{i+1}f_{i+2}.$ It commutes with $F_{i+l}$ for all $l$ and hence the unitary $R$-matrix constructed out of it solves the $(d, 3,l)$-gYBE for an $\cS^d_1$ realization of the supercharges. In a similar manner the other supercharges used to generate GHZ-class states give similar $F_i$ projectors that solve the $(d,3,l)$-gYBE in a unitary way. 

\paragraph{W-states}
By now the algorithm is clear. In what follows, we write down just the answers for the appropriate supercharges. To generate an entangled state which is composed by the same basis as the standard W-state $\left[\ket{001}+\ket{010}+\ket{100}\right]/\sqrt{3}$, the supercharge is 
\begin{equation}\label{uq4}
Q_i = \frac{1}{\sqrt{\alpha^2+\beta^2+\gamma^2}}\left[\alpha b_ib_{i+1}q_{i+2}^\dag + \beta b_i q_{i+1}^\dag b_{i+2} + \gamma q_i^\dag b_{i+1} b_{i+2}\right],
\end{equation}
which generates the projector to the bosonic sector
\begin{eqnarray}
B_i & = & \frac{1}{\alpha^2 + \beta^2 + \gamma^2}\left[\alpha^2 b_ib_{i+1}f_{i+2} + \alpha\beta b_iq_{i+1}q_{i+2}^\dag + \alpha\gamma q_ib_{i+1}q_{i+2}^\dag  + \alpha\beta b_iq_{i+1}^\dag q_{i+2}\right. \nonumber \\
&  &+ \left.  \beta^2 b_if_{i+1}b_{i+2} + \beta\gamma q_iq_{i+1}^\dag b_{i+2} +\alpha\gamma q_i^\dag b_{i+1} q_{i+2} + \gamma\beta q_i^\dag q_{i+1}b_{i+2} + \gamma^2 f_ib_{i+1} b_{i+2} \right].
\end{eqnarray}
This commutes with $B_{i+l}$ for $l\geq 3$ thus giving a unitary $R$-matrix of the form \eqref{Rcp} that solves the $(d,3,l)$-gYBE for $l\geq 3$ with an $\cS^d_1$ realization of the supercharge. 
Then, 
\begin{eqnarray}
R_i(u)\ket{001} & = &  \frac{1}{\alpha^2+\beta^2+\gamma^2}\left[\left(\alpha^2e^{cu}+\beta^2+\gamma^2\right)\ket{001} +  \left(e^{cu}-1\right)\left\{\alpha\beta\ket{010}+\alpha\gamma\ket{100}\right\}\right], 
\nonumber \\
R_i(u)\ket{010} & = & \frac{1}{\alpha^2+\beta^2+\gamma^2}\left[\left(\alpha^2+\beta^2e^{cu}+\gamma^2 \right)\ket{010}  +\left(e^{cu}-1\right)\left\{\alpha\beta \ket{001} + \beta\gamma\ket{100}\right\}\right],
 \nonumber \\
R_i(u)\ket{100} & = &  \frac{1}{\alpha^2+\beta^2+\gamma^2}\left[\left(\alpha^2+\beta^2+\gamma^2 e^{cu}\right)\ket{100} +  \left(e^{cu}-1\right)\left\{\alpha\gamma\ket{001}+\beta\gamma\ket{010}\right\}\right] \nonumber \\
\end{eqnarray}
for $d=2$. 
The supercharge in \eqref{uq4} generates a fermionic sector spanned by $\ket{000}$ with the corresponding projector, $F_i = Q_i^\dag Q_i = b_ib_{i+1}b_{i+2}$. The remaining product states are zero-modes. We could have equally constructed this bosonic projector from supercharges that generate other fermionic sectors spanned by either of the states, $\ket{011}, \ket{101}, \ket{110}, \ket{111}$. The construction for such relevant supercharges proceeds in a canonical fashion as outlined in all the previous examples, so we do not elaborate it further. 

For any other state of the W-states, we can do similarly by choosing a relevant supercharge. For example, entangled states composed by the same basis as the standard W-state $\left[\ket{111}+\ket{001}+\ket{010}\right]/\sqrt{3}$ can be constructed from
\begin{equation}\label{uq5}
Q_i = \frac{1}{\sqrt{\alpha^2+\beta^2+\gamma^2}}\left[\alpha q_i^\dag q_{i+1}^\dag q_{i+2}^\dag + \beta b_i b_{i+1} q_{i+2}^\dag + \gamma b_i q_{i+1}^\dag  b_{i+2}\right],
\end{equation}
which generates
\begin{eqnarray}
B_i & = & \frac{1}{\alpha^2 + \beta^2 + \gamma^2}\left[\alpha^2 f_if_{i+1}f_{i+2} + \alpha\beta q_i^\dag q_{i+1}^\dag f_{i+2} + \alpha\gamma q_i^\dag f_{i+1}q_{i+2}^\dag + \alpha\beta q_iq_{i+1} f_{i+2}\right. \nonumber \\
&  &+ \left. \beta^2 b_ib_{i+1}f_{i+2} + \beta\gamma b_iq_{i+1} q_{i+2}^\dag  +\alpha\gamma q_i f_{i+1} q_{i+2} + \gamma\beta b_i q_{i+1}^\dag q_{i+2} + \gamma^2 b_if_{i+1} b_{i+2} \right]. 
\end{eqnarray}

\paragraph{Partially entangled states}
Consider the partially entangled state in the $A-BC$ class given by $\left[\ket{000}+\ket{011}\right]/\sqrt{2}$. States spanned by the same basis are created by the supercharge 
\begin{equation}\label{uq6}
Q_i = \frac{1}{\sqrt{\alpha^2+\beta^2}}\left[\alpha q_i q_{i+1} q_{i+2} + \beta q_i f_{i+1} f_{i+2} \right]
\end{equation}
from $\ket{111}$, 
which creates the projector 
\begin{equation}\label{b6}
B_i = \frac{1}{\alpha^2+\beta^2}\left[\alpha^2 b_ib_{i+1}b_{i+2} + \alpha\beta\left(b_i q_{i+1}q_{i+2} + b_i q_{i+1}^\dag q_{i+2}^\dag\right) + \beta^2 b_if_{i+1}f_{i+2}\right].
\end{equation}
As before, this projector commutes with $B_{i+l}$ for all $l\geq 3$ giving a unitary $R$-matrix that solves the $(d,3,l)$-gYBE for all $l\geq 3$ with an $\cS^d_1$ realization of the supercharge. 
For $d=2$, we end up with
 \begin{eqnarray}
R_i(u)\ket{000} & = &  \frac{1}{\alpha^2+\beta^2}\left[\left(\alpha^2e^{cu}+\beta^2\right)\ket{000} + \alpha\beta \left(e^{cu}-1\right)\ket{011}\right], \nonumber \\
R_i(u)\ket{011} & = & \frac{1}{\alpha^2+\beta^2}\left[\alpha\beta \left(e^{cu}-1\right)\ket{000} + \left(\alpha^2+\beta^2e^{cu}\right)\ket{011} \right].
\end{eqnarray}

Similarly, a representative state of the partially entangled class $AC-B$ is generated by a bosonic projector constructed from 
\begin{equation}\label{uq7}
Q_i = \frac{1}{\sqrt{\alpha^2+\beta^2}}\left[\alpha q_i q_{i+1} q_{i+2} + \beta f_i q_{i+1} f_{i+2} \right],
\end{equation}
while a state in the partially entangled class $AB-C$  
is generated by a bosonic projector built out of 
\begin{equation}\label{uq8}
Q_i = \frac{1}{\sqrt{\alpha^2+\beta^2}}\left[\alpha q_i q_{i+1} q_{i+2} + \beta f_i f_{i+1} q_{i+2} \right].
\end{equation}
These exhaust all the SLOCC classes for a three-qubit system. 

\subsubsection*{General $m$}

We write down supercharges to get the bosonic projectors for just the $m$-qubit GHZ state class and the $m$-qubit W-state class.

States composed by the same basis as the standard $m$-qubit GHZ state $\left[\ket{00\cdots0} + \ket{11\cdots 1}\right]/\sqrt{2}$ are generated by the unitary $R$-matrix constructed using
\begin{equation}
Q_i = \frac{1}{\sqrt{\alpha^2+\beta^2}}\left[\alpha b_i \prod_{j=1}^{m-1}~q_{i+j}  + \beta q_i^\dag\prod_{j=1}^{m-1} f_{i+j}  \right],
\end{equation} 
which results in the projector to the bosonic sector 
\begin{equation}\label{b8}
B_i = \frac{1}{\alpha^2+\beta^2}\left[\alpha^2 \prod_{j=0}^{m-1}~b_{i+j} + \alpha\beta\left(\prod_{j=0}^{m-1}~q_{i+j} + \prod_{j=0}^{m-1}~q_{i+j}^\dag \right) + \beta^2 \prod_{j=0}^{m-1}~f_{i+j}\right].
\end{equation}
This projector commutes with $B_{i+l}$ for all $l\geq m$ and thus helps construct unitary $R$-matrices that satisfy
 the $(d,m,l)$-gYBE for all $l\geq m$. 
Finally, the $d=2$ case gives 
\begin{eqnarray}
R_i(u)\ket{00\cdots 0} & = &  \frac{1}{\alpha^2+\beta^2}\left[\left(\alpha^2e^{cu}+\beta^2\right)\ket{00\cdots 0} + \alpha\beta \left(e^{cu}-1\right)\ket{11\cdots 1}\right], \nonumber \\
R_i(u)\ket{11\cdots 1} & = & \frac{1}{\alpha^2+\beta^2}\left[\alpha\beta \left(e^{cu}-1\right)\ket{00\cdots0} + \left(\alpha^2+\beta^2e^{cu}\right)\ket{11\cdots 1} \right].
\end{eqnarray}
For the other standard $m$-qubit GHZ states, similar entangled states can be generated from appropriate supercharges. 

As for the standard $m$-qubit W-state $\sum_{r=1}^m~\ket{0\cdots 01_r0\cdots 0}/\sqrt{m}$, we consider  the unitary $R$-matrix constructed out of
\begin{equation}\label{uq9}
Q_i = \frac{1}{\sqrt{\sum_{p=1}^m~\alpha_p^2}}\sum_{r=0}^{m-1}~\alpha_{r+1}\left(\prod_{j=0}^{r-1}~b_{i+j}\right)q_{i+r}^\dag\left(\prod_{j=r+1}^{m-1}~b_{i+j}\right).
\end{equation}
This generates the projector $B_i=Q_iQ_i^\dag$ which commutes with $B_{i+l}$ for all $l\geq m$. The $B_i$ in turn is used to construct the unitary $R$-matrix that solves the $(d,m,l)$-gYBE for all $l\geq m$ 
with the final result for $d=2$ 
\begin{eqnarray}
R_i(u)\ket{0\cdots 01_r0\cdots 0} & = & \frac{1}{\sum_{p=1}^m\alpha_p^2}\left[\left(\alpha_r^2e^{cu}+\sum_{s(\neq r)} \alpha_s^2\right)\ket{0\cdots 01_r0\cdots 0} \right. \nonumber \\
 & & \hspace{17mm} \left.+\sum_{s(\neq r)}\alpha_r\alpha_s\left(e^{cu}-1\right)\ket{0\cdots 01_s0\cdots 0}\right].
 \end{eqnarray}

\subsubsection*{Unitary solutions for $1\leq l<m$}

The unitary solutions constructed above obey the $(d,m,l)$-gYBE for $l\geq m$. Here we present other unitary solutions which solve the gYBE for $1\leq l<m$. It turns out that the solutions with the $\cS^2_1$ realization generate only the product states, whereas $\cS^d_1$ ($d>2$) realizations lead to entangled states. For simplicity, we only consider the $d=3$ case for higher $\cS^d_1$ realizations. 

Let us start by considering the case of  $m=2, l=1$ and the supercharge 
\begin{equation}
Q_i = q_iq_{i+1},
\end{equation}
generating the Hamiltonian
\begin{equation}
H_i = \{Q_i, Q_i^\dag\}= b_ib_{i+1} + f_if_{i+1}.
\end{equation}
This is a commuting projector with $\left[H_i, H_{i+1}\right]=0$ and can then be used to construct a unitary $R$-matrix of the form 
\begin{equation}\label{moreunitaryR}
R_i(u)=I + (e^{cu}-1)H_i.
\end{equation} 
Note that this Hamiltonian is left invariant by $q_iq_{i+1}^\dag$ and $q_i^\dag q_{i+1}$. When supersymmetry is realized using $\cS^2_1$, one can see that the Hamiltonian acts on the states as 
\begin{equation}
H_i\ket{00}=\ket{00},\qquad H_i\ket{11}=\ket{11},\qquad H_i\ket{01}=H_i\ket{10}=0.
\end{equation}
Clearly the $R$-matrix maps products states to product states. If one considers the $\cS^3_1$ realization, the Hamiltonian still maps product states to product states:
\begin{eqnarray}
 & & H_i\ket{11}=H_i\ket{22}=H_i\ket{12}=H_i\ket{21}=\frac12\ket{\tilde{1}\tilde{1}}, \qquad H_i\ket{00}=\ket{00},  \nonumber \\
 & & H_i\ket{01}=H_i\ket{02}=H_i\ket{10}=H_i\ket{10}=0,
\end{eqnarray} 
where $\ket{\tilde{1}}=\left[\ket{1}+\ket{2}\right]/\sqrt{2}$. However, the $R$-matrix generates entangled states as 
\begin{equation}
R_i(u)\ket{ab}=\ket{ab}+\frac12\left(e^{cu}-1\right)\ket{\tilde{1}\tilde{1}}, \qquad a,b=1,2.
\label{moreunitary1}
\end{equation}
Here, we can effectively consider (\ref{moreunitary1}) as a qubit system, since $\ket{0}$ does not appear. Then, it can be regarded as the same SLOCC class of the Bell states. 
For example, $\ket{11}$ and $\ket{\tilde{1}\tilde{1}}$ in the case of $a=b=1$ are mapped to $\ket{11}$ and $\frac{1}{\sqrt{2}}\ket{22}$ respectively, by the ILO $\begin{pmatrix}1 & -1 \\ 0 & 1\end{pmatrix}^{\otimes 2}$.   
The supercharge $Q_i=q_i^\dag q_{i+1}^\dag$ gives the same Hamiltonian leading to the same result, whereas the supercharges $Q_i=q_iq_{i+1}^\dag$ (equivalently $q_i^\dag q_{i+1}$) do not generate 
entangled states even for $\cS^3_1$, ending up with the result
\begin{equation}
R_i(u)\ket{0a}=\ket{0}\left\{\ket{a}+\frac{1}{\sqrt{2}}\left(e^{cu}-1\right)\ket{\tilde{1}}\right\},
\qquad R_i(u)\ket{a0}=\left\{\ket{a}+\frac{1}{\sqrt{2}}\left(e^{cu}-1\right)\ket{\tilde{1}}\right\}\ket{0}.
\end{equation}

The next simple case is $m=3, l=1, 2$, with the associated supercharge
\begin{equation}
Q_i = q_iq_{i+1}q_{i+2},
\end{equation}
generating the Hamiltonian
\begin{equation}
H_i = b_ib_{i+1}b_{i+2} + f_if_{i+1}f_{i+2}.
\end{equation}
It is easy to check that these are once again commuting projectors with $\left[H_i, H_{i+1}\right]=0$ as well as $\left[H_i, H_{i+2}\right]=0$, so that the construction above applies to $l=1,2$. 
The $R$-matrix of the same form (\ref{moreunitaryR}) generates entangled states under the $\cS^3_1$ realization:
\begin{equation}
R_i(u)\ket{a_1a_2a_3}=\ket{a_1a_2a_3}+\frac{1}{2\sqrt{2}}\left(e^{cu}-1\right)\ket{\tilde{1}\tilde{1}\tilde{1}}, \qquad a_1,a_2,a_3=1,2,
\end{equation}
which can be regarded as the same SLOCC class as the GHZ states. 
As another example, the Hamiltonian generated by the supercharge 
\begin{equation}
Q_i=q_iq_{i+1}q_{i+2}^\dag
\end{equation}
is 
\begin{equation}
H_i=b_ib_{i+1}f_{i+2} + f_if_{i+1}b_{i+2},
\end{equation}
which satisfies the same properties as above and the $R$-matrix (\ref{moreunitaryR}) leads to partially entangled states under the $\cS^3_1$ realization:
\begin{equation}
R_i(u)\ket{a_1a_20}=\left\{\ket{a_1a_2}+\frac12\left(e^{cu}-1\right)\ket{\tilde{1}\tilde{1}}\right\}\ket{0}.
\end{equation}
Also, the $R$-matrix from the supercharge 
\begin{equation}
Q_i=q_iq_{i+1}w_{i+2}
\end{equation}
gives another type of partially entangled states with $\cS^3_1$:
\begin{equation}
R_i(u)\ket{a_1a_2a_3}=\left\{\ket{a_1a_2}+\frac12\left(e^{cu}-1\right)\ket{\tilde{1}\tilde{1}}\right\}\ket{a_3}.
\end{equation}

Now it is straightforward to generalize to arbitrary $m$ and $l<m$.
For example, let us consider the supercharge
\begin{equation}
Q_i = \prod_{j=0}^{m-1}~q_{i+j},
\end{equation}
generating the Hamiltonian
\begin{equation}
H_i = \prod_{j=0}^{m-1}~b_{i+j} + \prod_{j=0}^{m-1}~f_{i+j}.
\end{equation}
Once again these are projectors and satisfy $\left[H_i, H_{i+l}\right]=0$ for all $l<m$, thereby solving the $(d,m,l)$-gYBE for all $l<m$. 
Under the $\cS^3_1$ realization, the $R$-matrix of the form (\ref{moreunitaryR}) generates entangled states as 
\begin{equation}
R_i(u)\ket{a_1\cdots a_m}=\ket{a_1\cdots a_m} + \frac{1}{2^{m/2}}\left(e^{cu}-1\right)\ket{\tilde{1}\cdots\tilde{1}},
\end{equation}
with $a_1,\ldots, a_m=1,2$. 

It seems non-trivial to generate entangled states falling in the class of the W-states in this manner. 

\subsection{The Rowell-Wang solutions from supersymmetry}
\label{susyRW}

The unitary $R$-matrix without a spectral parameter that produces the Bell states upon acting on the product basis of two qubits is the so-called {\it Bell matrix} given by 
\begin{equation}\label{bellm}
R_i = \frac{1}{\sqrt{2}}\left(\begin{array}{cccc}1 & 0 & 0 & 1\\ 0 & 1 & 1 & 0\\ 0 & -1 & 1 & 0\\ -1 & 0 & 0 & 1\end{array}\right)= \frac{1}{\sqrt{2}}\left[1+x_i\right],
\end{equation}
with 
\begin{equation}
\label{rwrep}
x_i = \textrm{i} \sigma^y_i\otimes \sigma^x_{i+1}
\end{equation}
being the generators of the {\it extraspecial 2-group} \cite{r1,r2}. They obey
\begin{equation}
\label{exsrelations}
x_i^2 = -1,\qquad x_ix_{i+1}=-x_{i+1}x_i,\qquad x_ix_j=x_jx_i,~\textrm{for}~ |i-j|>1.	
\end{equation}
It is easy to check that (\ref{bellm}) satisfies the braid relations \eqref{braid}.\footnote{In \cite{hietarinta,dye}, solutions of (\ref{rcheck}) with suppressed spectral-parameter dependence are found in some cases. Multiplying the solutions there by the permutation matrix will give solutions of the braid relations. In \cite{GKRZ}, representations of the braid group are investigated by using twisted tensor products.}  
We actually see that $R_iR_{i+1}R_i \propto x_i+x_{i+1}$. Baxterized forms of this solution will introduce spectral parameter dependence. In the literature there exist two different forms of the Baxterized version of these braid solutions: in \cite{r2}, via the Baxterization procedure, and in \cite{jonesBaxterization,mo}, as a {\it type II} solution of the YBE. As $R_i$ in (\ref{bellm}) satisfies $R_i^2=\sqrt{2}R_i-1$ and the braid relation, we see that it satisfies the YBE with a Baxterized form similar to the one in (\ref{Rcp}) as well. We discuss in Sec.~\ref{remarks} different Baxterized versions of the solutions obtained from supersymmetry and the related braid-like algebras. 

We can realize the extraspecial 2-group generators from supersymmetry by noticing that
\begin{equation}\label{xSUSY}
x_i = -w_i\left(q_i+q_i^\dag\right)\left(q_{i+1}+q_{i+1}^\dag\right)
\end{equation}
satisfies 
$x_i^2=-1$ and $x_ix_{i+1}=-x_{i+1}x_i$ for the $\cS^2_1$ realization (\ref{s21}). This is due to the fact that the Witten operator $w_i$ anticommutes with the supercharges $q_i$ and $q_i^\dag$. The far-commutativity is also trivially satisfied as these pairs of generators have trivial common support. One can easily check that $q+q^\dag=\sigma^x$ and $w=1-2qq^\dag = -\sigma^z$. This makes the $x_i$ in \eqref{xSUSY} precisely equal to the $x_i$ in \eqref{rwrep}.

Choosing $\cS^d_1$ ($d>2$) instead, the $x_i$ in \eqref{xSUSY} no longer satisfy $x_i^2=-1$, but are such that 
\begin{equation}
x_i^2 = -h_ih_{i+1},
\end{equation}
where $h_i$ and $h_{i+1}$ are local supersymmetric Hamiltonians, which are also projectors. 
The relation $x_ix_{i+1}=-x_{i+1}x_i$ continues to hold and one also has that $x_i^3=-x_i$, as can be easily verified using \eqref{sissusy}. In this case the operator 
\begin{equation}
b_i =\frac{1}{2}\left[h_ih_{i+1} + x_i\right]
\end{equation}
satisfies the braid relation, $b_ib_{i+1}b_i = b_{i+1}b_ib_{i+1}$, but it is not invertible. 

These arguments can be generalized to the multi-qubit case to produce the GHZ states by $R_i=\left[1+x_i\right]/\sqrt{2}$ with the extraspecial 2-group generators now given by
\begin{equation}\label{rwGHZm}
x_j= \textrm{i}\sigma^y_j\prod_{k=1}^{m-1}~\sigma^x_{j+k}.
\end{equation}
It is easy to verify that these generators satisfy a generalized version of (\ref{exsrelations}):
\begin{eqnarray}
& & x_i^2=-1, \nonumber \\
& & x_ix_{i+l} = -x_{i+l}x_i \qquad (l=1,\cdots,m-1), \nonumber \\
& & x_i x_j= x_j x_i \qquad (|i-j|>m-1). 
\end{eqnarray}
The expression for $x_i$ in \eqref{xSUSY} can be generalized to 
\begin{equation}
x_i = -w_i\left(q_i+q_i^\dag\right)\left[\prod_{j=1}^{m-1}~\left(q_{i+j}+q_{i+j}^\dag\right)\right],	
\end{equation}
which precisely matches \eqref{rwGHZm} when supersymmetry is realized using $\cS^2_1$. This shows that the unitary Rowell-Wang solutions can be easily obtained from supersymmetry, and are in fact a special case of our construction.

\subsection{General structure}
\label{mainresult}

After gathering intuition with the specific situations analyzed up to this point, it is easy to uncover the general structure underlying our construction. Supersymmetry grades the Hilbert space into bosonic, fermionic and zero-mode parts. The supercharges swap the bosonic and the fermionic sectors, naturally creating orthogonal projectors to these parts. To generate an entangled state from a product state, it is then sufficient to group the components of the entangled state into either the bosonic or fermionic sector (this choice is just a convention), and place the initial product state into the other sector. This grading can be realized in a canonical way with a  supercharge built from SISs. We then make the following assertion:

{\it For any entangled state with at most $2^m-1$ product-state basis elements in an $m$-qubit system, there exists a unitary $R$-matrix, built using the supersymmetry algebra realized from $\cS^2_1$ such that it solves the $(2,m,l)$-gYBE for appropriate $l$, and maps the product states that make up the chosen entangled state to an entangled state which is a superposition of the same product states, while leaving invariant the product states not occurring in the chosen entangled state. }

This is the main result of this paper. The supercharges constructed so far illustrate this for the standard form of the entangled states in different SLOCC classes. In order to show that this is actually more general, we illustrate the generation of $\left[\alpha\ket{00} + \beta\ket{11} + \gamma\ket{10}\right]/\sqrt{\alpha^2+\beta^2+\gamma^2}$ following this method. The supercharge that flips the product state $\ket{01}$ to this entangled state is given by
\begin{equation}
Q_i = \frac{1}{\sqrt{\alpha^2+\beta^2+\gamma^2}}\left[\alpha b_iq_{i+1} +\beta q_i^\dag f_{i+1} + \gamma q_i^\dag q_{i+1}\right],
\end{equation}
which generates the projector
\begin{eqnarray}
B_i  & = &\frac{1}{\alpha^2+\beta^2+\gamma^2}\left[ \alpha^2 b_ib_{i+1} + \alpha\beta q_iq_{i+1} + \alpha\gamma q_ib_{i+1}+ \alpha\beta  q_i^\dag q_{i+1}^\dag \right.\nonumber \\   
                              &  &\hskip 3cm + \left.  \beta^2 f_if_{i+1} + \beta\gamma f_iq_{i+1}^\dag +\alpha\gamma q_i^\dag b_{i+1} + \gamma\beta f_iq_{i+1} + \gamma^2 f_ib_{i+1}\right],
\end{eqnarray}
satisfying $B_iB_{i+l}B_i=B_{i+l}B_iB_{i+l}$ ($l\geq 2$). The unitary $R$-matrix built out of this projector projects the product states $\{\ket{00}, \ket{11}, \ket{10}\}$ into this entangled state. The fermionic projector $F_i=Q_i^\dag Q_i$ projects to the other product state, $\ket{01}$. This can be easily extended to the multi-qubit case. 

\section{Final remarks and outlook}
\label{remarks}

We have seen explicitly how supersymmetry provides a systematic framework to construct unitary and non-unitary $R$-matrices generating the entangled states of the different SLOCC classes of a multi-qubit system. The $R$-matrices we obtain depend, however, on a spectral parameter, obscuring a possible connection to braiding operators, which do not depend on such parameter. It would then be important to obtain the `unBaxterized' versions of our solutions.  
 
For the form of the YBE in \eqref{rbraid}, the limits $u\rightarrow 0, \infty$ should result in the braid generators. We could also consider a periodic function which would also result in the braid generators, as in \cite{baxterization2}. These considerations suggest that the braid generators corresponding to the $R$-matrices in Sec. \ref{susyR} are either trivial, or the nilpotent operators ($Q_i$), or the commuting projectors ($B_i$ and $H_i$). The latter, however, while satisfying the braid relations, are not invertible and thus the usual Baxterization of \cite{jonesBaxterization} does not apply here.

However, there could exist non-trivial braid-like algebras which accommodate non-invertible operators that can be Baxterized to satisfy the general form of the YBE in footnote~\ref{foot3}, as proposed in \cite{crampe1}. Their generators satisfy far-commutativity and
\begin{equation}
\label{nbraid}
\left[\sigma_{i+1}\sigma_i, \sigma_i+\sigma_{i+1}\right]=0,\qquad i=1,\ldots, n-2,
\end{equation}
instead of the usual braid relations, and need not be invertible. It was shown in \cite{crampe1} that a Baxterization of these generators using {\it two} spectral parameters,
\begin{equation}
\check{R}_i(x,y) = \left(1-y\sigma_i\right)\left(1-x\sigma_i\right)^{-1}, 
\end{equation} 
satisfies the YBE without the assumption of the difference property, namely
\begin{equation}
\label{nYBE}
\check{R}_i(x,y)\check{R}_{i+1}(x,z)\check{R}_i(y,z) = \check{R}_{i+1}(y,z)\check{R}_i(x,z)\check{R}_{i+1}(x,y).
\end{equation}

It is easy to check that the nilpotent operators ($Q_i$) constructed out of SISs and the corresponding commuting projectors ($H_i$ or $B_i$) satisfy the braid-like relations in \eqref{nbraid}. Using the Baxterization procedure presented above one obtains 
\begin{eqnarray}
\check{R}_i(x,y)  =  1 + (x-y) Q_i, \qquad
\check{R}_i(x,y)  =  1 + \left(\frac{x-y}{1-x}\right) H_i
\end{eqnarray}
as the corresponding $R$-matrices satisfying \eqref{nYBE}. These $R$-matrices continue to have the same entangling properties as the ones constructed in Sec. \ref{susyR}. However, it is still unclear how they are connected to the braid group and knots.  

It is also worthwhile to note that for the extraspecial 2-group generated by $m_i$ satisfying \eqref{exsrelations}, the $R$-matrix 
\begin{equation}
\check{R}_i(x,y)=1+\left(\frac{x-y}{2-x-y}\right)m_i
\end{equation}
can be derived as a solution of (\ref{nYBE}) through relations with the Hecke algebra \cite{crampe1}. Also, the extraspecial 2-group generators 
obey another braid-like algebra $\mathcal{A}_n(0,0,-2)$ introduced in \cite{crampe2}. The entangling properties of this $R$-matrix are similar to the ones of the Rowell-Wang solutions, despite the two spectral parameters.  

It would be certainly very interesting to understand these issues better, as they would likely help clarifying the connection between topological and quantum entanglement. 

We can extend the analysis of this paper to multi-qudit systems by using the so-called {\it para-supersymmetry} instead of supersymmetry. The difference is that para-supercharges satisfy $q^d=0$, where $d$ corresponds to the dimension of the local Hilbert space. We are also obtaining the results of this paper for a non-local realization  of supersymmetry using {\it partition algebras}. We will present these results in a forthcoming work.

Some more speculative outlook concerns the understanding of the SLOCC classification through the YBE and the gYBE. Can the $R$-matrices provide some way to carry out this classification by providing a better `order parameter'? In this regard, it would be interesting to investigate the role played by the Witten index, if any, in this classification. 

As a different kind of generalizations of the YBE, the Zamolodchikov tetrahedron equation \cite{Zam_tetra1,Zam_tetra2} can be regarded as a fundamental equation of integrable systems in $(2+1)$ dimensions. It would be interesting to try to apply our solution-generating technique to such higher-dimensional systems. 

Finally, in recent years there have been some works connecting entanglement entropy in Chern-Simons theory with knot and link invariants \cite{SSW,bal1,DSDRZJ,bal2,Melnikov:2018zfn}. It would be interesting to see how the $R$-matrices fit in these works and discuss the relation between topological and quantum entanglement in a physical setting. 

\subsection*{Acknowledgements}

We are happy to thank A. P. Balachandran, S. Jordan, H. Katsura, D. Melnikov and E. Rowell for comments on the manuscript. PP and FS are supported by the Institute for Basic Science in Korea (IBS-R024-Y1, IBS-R018-D1). DT is supported in part by the INFN grant {\it Gauge and String Theory (GAST)}. 

\appendix
\section{Supercharges relating non-trivial SLOCC classes}
\label{more2}

The supercharges constructed in Sec. \ref{susyR} swap two SLOCC classes, the product-state class and the entangled-state class. In the same spirit, it is possible to construct supercharges that swap two different entangled-state classes. Consider for example the three-qubit sector, with two non-trivial SLOCC classes describing tripartite entanglement, namely the GHZ state class and the W-state class. The supercharge that swaps these two is given by 
\begin{eqnarray}
&& Q_i  =
\alpha_1 b_ib_{i+1}q_{i+2}^\dag + \alpha_2 b_iq_{i+1}^\dag b_{i+2} + \alpha_3 q_i^\dag b_{i+1}b_{i+2} 
       +\alpha_1 q_iq_{i+1}f_{i+2} + \alpha_2 q_if_{i+1}q_{i+2} + \alpha_3 f_i q_{i+1}q_{i+2},\cr &&
\end{eqnarray}
with $\alpha_1,\alpha_2,\alpha_3$ being real coefficients. $B_i=Q_iQ_i^\dag$ ($F_i=Q_i^\dag Q_i$) is a projector to the bosonic (fermionic) sector satisfying 
\begin{equation}
B_i^2 = kB_i, \qquad F_i^2=kF_i ,\qquad k\equiv 2\left(\alpha_1^2+\alpha_2^2+\alpha^2_3\right),
\end{equation}
and generating the W-state (GHZ state) class. Using $H_i=B_i+F_i$, one can then construct a unitary $R$-matrix, as in Sec. \ref{susyR}, to obtain both inequivalent tripartite SLOCC classes of the three-qubit case. 
For example, 
\begin{eqnarray}
R_i(u)\ket{001} & = & \left\{1+\frac{2\alpha_1^2}{k}\left(e^{cu}-1\right)\right\}\ket{001} +\frac{2\alpha_1}{k}\left(e^{cu}-1\right)\left[\alpha_2\ket{010}+\alpha_3\ket{100}\right], \nonumber \\
R_i(u)\ket{000} & = & \frac12\left[\left(e^{cu}+1\right)\ket{000}+\left(e^{cu}-1\right)\ket{111}\right].
\end{eqnarray}

This procedure naturally generalizes to the multi-qubit sector. These arguments show the versatility of our method to construct unitary $R$-matrices that generate the different entangled states of a multi-qubit system.
 

\end{document}